\documentclass[a4paper,usenatbib,usegraphicx,useAMS,fleqn]{mn2e}
\usepackage{graphicx}
\usepackage{amsfonts,amsmath,amssymb,gensymb}
\usepackage{natbib}
\usepackage{aas_macros,apjfonts}
\usepackage{rotating,array,multirow,array,longtable,tabularx,color,bm}
\usepackage{pstricks,epsf, epsfig}
\usepackage{afterpage,fix2col}

\def\be{\begin{equation}}
\def\ee{\end{equation}}

\title [Spin alignment in BH binaries]
{Spin alignment and differential accretion \\ in merging black hole binaries}
\author[D.~Gerosa et al.]{
D.~Gerosa$^1 \thanks{E-mail: d.gerosa@damtp.cam.ac.uk}$,
B.~Veronesi$^2	 \thanks{E-mail: benedetta.veronesi@studenti.unimi.it}$, 
G.~Lodato$^2\thanks{E-mail: giuseppe.lodato@unimi.it}$
\&  G.~Rosotti$^3\thanks{E-mail: rosotti@ast.cam.ac.uk}$\\
$^1$ Department of Applied Mathematics and Theoretical Physics, Centre for Mathematical Sciences, University of Cambridge, \\ Wilberforce Road, Cambridge CB3 0WA, UK\\
$^2$ Dipartimento di Fisica, Universit\`a degli studi di Milano, Via Celoria 16, Milano, I-20133, Italy\\
$^3$ Institute of Astronomy, University of Cambridge, Madingley Road, Cambridge CB3 0HA, UK
}

\pubyear{2015}

\begin{document}
\maketitle

\begin{abstract}
Interactions between a supermassive black hole binary and the surrounding accretion disc can both assist the binary inspiral and align the black hole spins to the disc angular momentum.  While binary migration is due to angular-momentum transfer within the circumbinary disc, the spin-alignment process is driven by the mass accreting on to each black hole. Mass transfer between different disc components thus couples the inspiral and the alignment process together. Mass is expected to leak through the cavity cleared by the binary, and preferentially accretes on to the lighter (secondary) black hole which orbits closer to the disc edge. Low accretion rate on to the heavier (primary) black hole slows the alignment process down.  We revisit the problem and develop 
a semi-analytical model to describe the coupling between gas-driven inspiral and spin alignment, finding that binaries with mass ratio $q\lesssim 0.2$ approach the gravitational-wave driven inspiral in differential misalignment: light secondaries prevent primaries from aligning. Binary black holes with misaligned primaries are ideal candidates for precession effects in the strong-gravity regime and may suffer from moderately large ($\sim 1500$ km/s) recoil velocities.
\end{abstract}


\section{Introduction}

 Following a galaxy merger, the supermassive black holes (BHs) hosted \citep{kormendy95} by the two merging galaxies sink towards the centre of the newly formed stellar environment through dynamical friction, forming a binary \citep{begelman80,mayer13}. Binary BHs can merge if the astrophysical environment provides a way to dissipate their angular momentum in less than a Hubble time. Scattering from stars can bring the binary only down to parsec scales \citep{frankrees76}, below which the available phase space is quickly depleted, thus stalling the inspiral process (\emph{final-parsec problem}, \citealt{milos01,yu02}). While triaxiality in the stellar potential may help driving the inspiral in elliptical gas-poor galaxies \citep{merrittpoon04,Berczik06}, the interaction with gaseous disc(s) may actually solve the final-parsec problem in gas-rich galaxies \citep{armitagenararajan02,cuadra09}. Indeed, dynamical friction against a gaseous background can promote the merger bringing the binary separation to distances of the order of 0.1 pc within a time-scale of 10-50 Myrs \citep{escala05,dotti07,dotti09b}. Further shrinking of the binary can proceed through what is known as type II migration in the context of planet-disc interaction. However, such disc-assisted migration can only be effective at separations smaller than $\sim$ 0.01 pc, beyond which the disc becomes self-gravitating and will likely fragment and form stars \citep{lodato09}. Finally, if the binary reaches separations close to $\simeq 10^{-3} {\rm pc}$, gravitational-waves  (GWs) quickly become an extremely efficient way to drive the binary to a prompt merger \citep{petmat63,peters64}. Asymmetric emission of GWs in the late inspiral and merger may imprint recoil velocities to the remnant BHs \citep{redmount89} which can be as high as $5000$ km/s \citep{campanelli2007b,gonzalez2007,lousto11}, possibly larger than the escape speed of the host galaxies \citep{favata04}.

A variety of electromagnetic signatures has been proposed to detect supermassive BH binaries, which however remain elusive \citep{dotti12,schnittman13,bog15}. The most convincing evidence comes from double active galactic nuclei (AGN) imaging, with the notable example of the radio galaxy 0402+379 showing two compact cores with estimated separation of $7.3$ pc \citep{rodriguez06}. Indirect evidence for supermassive BH binaries at sub-parsec scales suffer from higher uncertainties; we mention in particular the case of the blazar OJ287 \citep{valtonen08}, where  $\sim 12$ yr periodic outbursts have been interpreted as signature of a BH binary orbital motion. Identifying recoiling BHs trough observations is even more challenging \citep{komossa12}, but tentative candidates are nonetheless present  \citep{komossa08,civano12,Decarli14,koss14}. 
Direct measurements of supermassive BH inspirals and mergers are the main target of all future space-based GW observatories. The eLISA (evolved Laser Interferometer Space Antenna) mission, recently approved by the European Space Agency, is expected to detect hundreds merging binaries per year casting new light on our understating of such systems  \citep{eLISA,BHeLISA}.

Hydrodynamical interactions can not only assist the binary inspiral, but  are responsible for the reorientation of the two BH spins. BH spin alignment has a crucial impact on the merger dynamics and on the cosmic growth history of supermassive BHs. Strong recoil velocities can only be achieved if the merging BHs are highly spinning and the two spin vectors are strongly misaligned between each other. Highly recoiling BHs can be significantly displaced from the galactic nucleus or even ejected from it. This has strong consequences on the coevolution of BHs and their host galaxy, with mild recoil regulating the BH growth and large kicks  velocities strongly affecting the feedback process \citep{blecha08}. If large recoils causing BH ejections are present, this affects the fraction of galaxies hosting supermassive BHs \citep{schnittman07,volonteri10,gerosa15} and consequently the predicted (e)LISA event rates \citep{sesana07b}.
From the GW data analysis point of view, spin misalignments introduce a richer structure in the expected signals that  carries precious information on the binary dynamics \citep{vecchio04,klein09,OShaughnessy13} and can improve the parameter estimation process by up to an order of magnitude \citep{chat14}. At the same time, since accurate waveform modelling is required in GW searches, spin precession makes the waveform generation more challenging \citep{Pekowsky,klein14}, dramatically increasing the parameter-space dimensions that need to be explored.

It is thus important for both electromagnetic and GW observations to understand in which region of the BH-binary parameter space we expect  significant spin misalignment, which is the goal of this paper.
The physical process responsible for the reorientation of BH spins during the long gas-driven inspiral is the so-called Bardeen-Petterson effect  \citep{bardeen75}, where the general relativistic Lense-Thirring torque between the BH and a misaligned disc  warps the accretion disc and secularly aligns the BH spin with the disc angular momentum. The Bardeen-Petterson effect does not just affect the binary dynamics during the gas-driven phase, but leaves a deep imprint in the subsequent GW-driven inspiral, where precession effects are strongly dependent on the residual misalignments left by the astrophysical environment \citep{kesden2010a,kesden2010b,gerosa13}. It is therefore an essential ingredient to predict the spin configuration at merger. 
 
The effectiveness of the Bardeen-Petterson effect in aligning the spins to the binary angular momentum within the time-scale of the merger has been recently investigated by multiple authors. \citet{bogdanovic07} first made some order-of-magnitude estimates of the alignment time for a single BH with its own disc and found that it is much smaller than the merger time concluding that alignment is likely in a gaseous environment. 
A similar conclusion was obtained by \citet{dotti10}, who found short alignment time-scales of $\sim 2 \times 10^6$ yr. One notable achievement of the study performed by \citet{dotti10} is combination of  smoothed-particle hydrodynamics (SPH) simulations and a semi-analytical treatment of the Bardeen-Petterson effect \citep{perego09}, through which they have been able to quantify the residual misalignment to either $10^\circ$ or $30^\circ$ depending on the gas temperature. In a previous work \citep{LG13}, some of us revisited these estimates considering the previously neglected effects of non-linear warps in the misalignment propagation trough the disc. Their conclusion is that the alignment time can be significantly longer than $10^7$ yr if the initial misalignment are large, thus casting doubts on the ability of the disc to align the binary. \citet{miller13} made a sensible step forward, pointing out that the spin alignment process in BH binaries may actually be sensibly faster than for isolated BHs, because of the stabilizing effect of the companion that increases the degree of disc warping  close to the holes.

In this paper, we argue that the binary mass ratio plays a key role in estimating the spin-alignment likelihood in merging BH binaries. We present a semi-analytical model to compute the inspiral and alignment processes from the properties of the circumbinary disc. On the one hand, the mass ratio strongly affects the binary inspiral rate, marking the onset of different disc-morphology regimes when either a gap or a proper cavity can be opened. On the other hand, and perhaps most importantly, the binary mass ratio sets the amount of differential accretion on to the two components of the binary system. The Bardeen-Peterson effectiveness in aligning the spins depends sensitively on the mass accretion rate through each single disc. We quantify this quantity constructing prescriptions based on results of hydrodynamical simulations. Gas is expected to preferentially accrete on to the lighter binary member that therefore aligns faster. Accretion rates on to the heavier BH is consequently smaller and may prevent it from aligning.

This paper is organized as follows. Our  model to compute the relevant time-scales is introduced in Sec.~\ref{discmodel}, where both the inspiral and the alignment problems are treated. 
Sec.~\ref{results} describes our main findings, namely the role of the binary mass ratio in the spin-alignment process, and its relevance  for cosmologically motivated   binary distributions and kick predictions. Finally, in Sec.~\ref{discussion}  we draw our conclusions and stress the possible caveats of our analysis.

\section{Binary and discs modelling}
\label{discmodel}
We model the gas environment surrounding  merging BH binaries through three different accretion discs: mass may reach the binary from galactic scales forming a \emph{circumbinary} disc, and later be accreted on to the individual BHs from \emph{circumprimary} and \emph{circumsecondary} discs respectively\footnote{The name \emph{minidiscs} can also be found in the literature to indicate circumprimary and circumsecondary discs.}.
We define $R$ to be the binary separation, $M_1$ and $M_2$ to be the two BH masses (with $M_1\geq M_2$), $M_{\rm bin}=M_1 + M_2$ to be the total mass of the binary, $q=M_2/M_1 
\leq 1$ to be the binary mass ratio and $S_{i}= a_i G M_i^2/c$ to be the spin of any of two BHs (where $i=1,2$ and dimensionless spin $0\leq a_i \leq1$). Following \citet{clarkesyer95}, we also define a measure of the local\footnote{For typical values of the surface density exponent, this is a reasonable estimate of the rigorous value obtained radially integrating the surface density \citep{shakura73}. In addition, the migration rates are set by the properties of the disc in the vicinity of the planet, and the local nature of this parameter is thus more relevant than the total disc mass.} circumbinary-disc mass at a generic radius $r$ as $M(r)=4\pi \Sigma(r)r^2$ and finally $q_{\rm disc}(r)= M(r)/M_{\rm bin}$ to be the disc-to-binary mass ratio. When studying the spin alignment, we will refer to the mass of the aligning BH as $M$ (meaning either $M_1$ or $M_2$) and the mass of the other binary member as $M_{\rm c}$.
While the accretion rate of the circumbinary disc $\dot M_{\rm bin}$ 
determines the inspiral process, the alignment time-scales are only  determined by the rates $\dot M_1$ and $\dot M_2$ at which mass reaches the circumprimary and circumsecondary discs respectively. It is also useful to define $f$ to be the  dimensionless value of $\dot M_{\rm bin}$ in terms of the Eddington accretion rate
\begin{align}
\dot M_{\rm bin}  = f  \frac{M_{\rm bin}}{ t_{\rm Edd}}\;,
\label{deff}
\end{align}
where $t_{\rm Edd}=\kappa_{\rm e} c / 4\pi G \simeq 4.5 \times 10^8 {\rm yr}$ is the \cite{salpeter64} time, and $\kappa_{\rm e}$ is the 
opacity for Thomson electron scattering.

In this section, we first present a new estimate for the inspiral time-scale by interpolating estimates computed in different regimes and we discuss the circumbinary disc self-gravity condition to evaluate such inspiral time-scale. Secondly, we summarize the main findings of \cite{LG13} on the spin-alignment time-scale and we explore the effect of the companion on the individual-discs structure. We finally model mass transfer and differential accretion on to the different discs. 

\subsection{Gas-driven inspiral}
\label{gasdriventime}
If BH mergers do happen in nature, 
it is likely that the gas-driven phase is the bottleneck of the whole binary evolutionary track. Therefore, the time spent by the binary in such phase gives us an estimate of the total time available to align the BH spins through environmental processes before merger.
Although gas-driven inspiral is mediated by the torques exerted by the disc on to the binary, a detailed description of the torques is not necessary to correctly derive the migration rates. Ultimately, the migration rate is controlled by the rate at which the disc is able to redistribute the angular momentum gained from the binary, and the torques will adjust to give the correct rate (e.g., \citealt{armitage10}). This same mechanism is called type II migration in the context of protoplanetary discs \citep{linpap86b}. Depending on the ratio between the BHs and the circumbinary disc masses, we identify three possible regimes:

\begin{enumerate}
\item For small mass ratios ($ M_2 \ll M(R) \ll M_1$), the secondary BH perturbs the disc of the primary, which reacts opening a gap\footnote{Binaries with very small mass ratios ($q\lesssim 10^{-4}$, see \citealt{armitage10})
cannot open a gap; however, such low mass ratios are not expected to be relevant in the supermassive BH context.}  
at the binary separation \citep{linpap79a,lubow94}. Tidal interactions between  gas particles and the secondary BH transfer angular momentum to the disc, thus decreasing the binary separation. The secondary BH behaves like a fluid element in the disc, evolving at the viscous rate \citep{armitagenararajan02}
\begin{align}
t_{{\rm in}} = t_\nu (R) \simeq \frac{R^2}{\nu}\;,
\label{tnu}
\end{align}
where  $\nu = \alpha c_s H$ is kinematic viscosity coefficient of the disc, 
usually \citep{shakura73}  rescaled to a dimensionless coefficient $\alpha$ with the speed of sound $c_s$  and  the disc height  $H$. 

\item If the secondary BH mass becomes comparable to the disc mass ($M_2\sim M(R) \ll M_1$), the disc cannot efficiently redistribute the momentum acquired from the binary. The shrinking rate consequently decreases. An analytical expression for the inspiral time-scale in this regime can be computed directly from the angular-momentum conservation equation in the thin-disc approximation \citep{clarkesyer95,ivanov99,lodato09,baruteau13} and reads
\begin{align}
t_{{\rm in}} = \frac{M_2 +M(R)}{M(R)}t_\nu (R) \;,
\label{type2}
\end{align}
which correctly reduces to $t_\nu (R)$ in the limit $M_2 \ll M(R)$.

\item For comparable mass binaries ($M_1\lesssim M_2$), the secondary-BH potential cannot be neglected. The gap at the secondary location now becomes a proper cavity in the disc with radius $\sim 2R$ cleared by both BHs \citep{macfadyen08}. A rich phenomenology may be present, including disc asymmetry and growing eccentricity, and can only be captured using hydrodynamical simulations   \citep{cuadra09,roedig11,roedig12,shi12}.  
An approximate expression for the inspiral time-scale in the comparable mass regime has been presented by \cite{rafikov12}, assuming the binary potential to be represented by a Newtonian potential produced by the binary total mass. He obtains
\begin{align}
t_{\rm in} = \frac{M_1 M_2}{M_{\rm bin}\; M(R)} t_{\nu} (R)\;,
\label{highq}
\end{align}
\end{enumerate}
where the correction factor $M_1/M_{\rm bin}$ models the expected speed up due to the higher angular momentum flux induced by the binary mass. The same mass-ratio dependence has been very recently obtained by  \cite{dottimerloni15}  integrating the torque at the edge of a $2R$-wide cavity

Here, we propose a smooth analytical interpolation between the time-scales obtained 
in the different regimes  given by
\begin{align}
t_{\rm in} = \frac{M_1}{M_{\rm bin}} \frac{M_2+M(R)}{M(R)} t_{\nu} (R)\;,
\label{highq1}
\end{align}
which correctly reduces to either Eq.~(\ref{tnu}), (\ref{type2}) or  (\ref{highq})  in the relevant limits. 
Various numerical factors in Eq. ~(\ref{highq}) --as already acknowledged by \cite{rafikov12}  himself-- and different possible definitions of the viscous time-scale may modify our this estimate of a factor $\sim$few.
The accretion rate of  the circumbinary disc $\dot M_{\rm bin}$ only enters in the merger time-scale through the viscous time-scale, which can be rewritten as
\begin{align}
\label{tnudef}
t_\nu(R) = \frac{3}{4} \frac{M(R)}{\dot M_{\rm bin}} \;,
\end{align} 
since both $M(R) =  4\pi R^2 \Sigma(R)$ and ${\dot M_{\rm bin}} \simeq 3 \pi \nu \Sigma(R)$ are  related to the surface density $\Sigma$ of the circumbinary disc in the stationary limit.
Combining Eqs.~(\ref{deff}), (\ref{highq1}) and (\ref{tnudef}), we obtain our final estimate of the inspiral time-scale, to be compared to the individual alignment time-scales, 
\begin{align}
t_{\rm in}= \frac{3}{4} \frac{(1+q)\: q_{\rm disc}(R) +q}{(1+q)^2} \;  
\frac{t_{\rm{edd}}}{f} \;.
\label{tmergeq}
\end{align}

While the low mass ratio regime is relatively well tested in the planetary community (e.g. \citealt{Nelson2000Migration,Bate2003}, but see below for possible caveats), the regime of high mass ratios has not been explored extensively. At the moment only few simulations of disc driven migration of a binary have been conducted \citep{macfadyen08,cuadra09,shi12}, which test only a small part of the parameter space. For example, for a ratio $q=1/3$, \citet{cuadra09} find in 3D  SPH simulations a migration rate of $\dot{R}/R = -2 \times 10^{-5} \ \Omega$, where $\Omega=(GM_{\rm bin}/R^3)^{1/2}$ is the orbital frequency of the binary. They compare this value with the analytical formula from \citet{ivanov99}, our Eq.~(\ref{type2}), which yields $\dot{R}/R = -3 \times 10^{-5} \ \Omega$, in very good agreement. \citet{macfadyen08} find in 2D grid-based simulations of $q=1$ binaries that the inspiral time-scale is roughly the viscous time-scale, reduced by 
$q_{\rm disc}$, which is consistent with Eq.~(\ref{tmergeq}).
The simulations described so far neglect the details of the angular momentum redistribution mechanism, which in the standard picture is the magneto-rotational instability (MRI, \citealt{balbus91}), and typically adopt the \citet{shakura73} $\alpha$ parametrization (e.g., \citealt{macfadyen08}) in order to reduce the computational cost. Only recently \cite{shi12} were able to perform global numerical simulations of migrating binaries including the MRI. They found that magnetohydrodynamics effects slightly enhance the migration rate with respect to the purely hydrodynamical case (a factor of  $\sim 3$ when compared to \citealt{macfadyen08}). They also observed that the accretion of material with a higher specific angular momentum than the binary can make the binary \textit{gain} angular momentum, which however is offset by the higher torques they measure from the disc. Given the number of other uncertainties present in the model, we are thus satisfied that our expressions can be used reliably.

Recent numerical simulations \citep{duffell14,duermann15} in the planetary community have questioned the validity of type II migration, casting doubts that a regime where the satellite behaves like a test particle exists at all. In particular, the simulations show that it is possible to achieve faster (up to a factor of 5) migration rates than what expected from Type II theory. These simulations have only been run  for a fraction of a viscous time, and it is still unclear if this result holds on the time-scale of the merger. For this reason we neglect these results in what follows, and note that this makes our estimates an upper limit for the merger time-scales.

Finally, we note that the simulations conducted so far, to the best of our knowledge, have explored relatively thick discs, with an aspect ratio ranging from 0.05 to 0.1 \citep[e.g.,][]{macfadyen08,cuadra09,roedig12,shi12}. This is significantly thicker than the value we derive in the next section and it is not clear how the results would change with more realistic values (cf. Sec.~\ref{discussion}).

\subsection{Self-gravity condition}
\label{fragmentation}

The inspiral time-scale reported in Eq.~(\ref{tmergeq}) depends on the binary separation $R$. 
For typical disc structures \citep{shakura73,goodman03}, $t_{\rm in}$ is a steep monotonically increasing function of $R$  \citep{haiman09}. Most time will be spent by the binary at large separations, while the remaining inspiral is completed rather quickly. The time available to align the spins --which the spin-alignment time must be compared to-- is roughly the inspiral time-scale $t_{\rm in}$ evaluated at the largest separation of the disc-driven evolution. 

A natural physical limit on the size of the circumbinary disc is set by the disc self gravity. Local  gravitational stability under axisymmetric disturbances 
 is guaranteed up to the fragmentation radius $R_{\rm f}$ where the \citeauthor{toomre64}'s (\citeyear{toomre64}) parameter equals unity:
\begin{align}
Q\equiv\frac{c_s \Omega}{\pi G \Sigma} =1\;.
\label{q=1}
\end{align}
At separation $R>R_{\rm f}$, self gravity cannot be neglected and the disc is gravitationally unstable (cf. e.g. \citealt{lodatoNC} for a review). The evolution of gravitationally unstable discs has been investigated in great details in recent years \citep{LR04,RLA05,CLC09}.
If the cooling time is smaller or of the order of the dynamical time \citep{gammie01}, the disc will fragment into gravitationally bound clumps, although the actual fragmentation threshold is debated \citep{meru12}. For values appropriate to AGN discs, the disc is expected to fragment, create stars and thus deplete the area surrounding the binary of gaseous material, possibly halting the inspiral \citep{lodato09}. 
Using the vertical-equilibrium equation $c_s/\Omega = H$, the self-gravity stability condition $Q=1$  can be rewritten as
\begin{align}
q_{\rm disc}(R)= \frac{M(R)}{M_{\rm bin}} \simeq 4 \frac{H}{R}\, ,
\label{qdisc}
\end{align}
evaluated at $R=R_{\rm f}$. The fragmentation radius is likely to lie in the outer region of the circumbinary disc,  dominated by gas-pressure and electron-scattering opacity \citep{shakura73}.  Assuming viscosity  to be proportional to the gas pressure ($\beta$-disc) and setting the mass of the accreting object to $M_{\rm bin}$, one gets for the fragmentation radius \citep{shakura73,goodman03,haiman09}
\begin{align}
\begin{aligned}
R_{\rm f} &\simeq 10^5 \frac{GM_{\rm bin}}{c^2}
  \left( \frac{M_{\rm bin}}{10^7 M_\odot}\right)^{-26/27}
 \left( \frac{f}{0.1}\right)^{-8/27}  
  \left(\frac{\alpha}{0.2}\right)^{14/27} 
  \\
  &\simeq 0.05
  \left( \frac{M_{\rm bin}}{10^7 M_\odot}\right)^{1/27}
 \left( \frac{f}{0.1}\right)^{-8/27}  
  \left(\frac{\alpha}{0.2}\right)^{14/27} ~{\rm pc}
  \,.
  \label{fragradius}
\end{aligned}
\end{align}
For a separation $r$ in such region, the disc aspect ratio reads
\begin{align}
\begin{split}
\frac{H}{r}  &= 0.001
 \left( \frac{r}{GM_{\rm bin}/c^2}\right)^{1/20}
 \left( \frac{M_{\rm bin}}{10^7 M_\odot}\right)^{-1/10} 
 \\
 &\times
  \left( \frac{f}{0.1}\right)^{1/5} 
  \left(\frac{\alpha}{0.2}\right)^{-1/10} \;.
 \label{hrregb}
  \end{split}
\end{align}

In this paper, we evaluate the inspiral time-scale  of Eq.~(\ref{tmergeq}) at the fragmentation radius: $R=R_{\rm f}$. This is a rather conservative assumption, being $t_{\rm in}$ monotonically increasing with $R$ \citep{haiman09} and being $R_{\rm f}$ the largest separation at which gas can be found under the form of a circumbinary disc.
We are therefore assuming --somehow overcoming the final parsec problem-- that some previous mechanisms
are efficient  enough to shrink the binary separation down to $R_{\rm f}$. 

  From Eqs.~(\ref{tmergeq}-\ref{hrregb}) we find that the inspiral time $t_{\rm in}$ scales only mildly with the viscosity $\alpha$ and the binary total mass $M_{\rm bin}$. As for the accretion rate $f$, the implicit dependence from Eq. (\ref{hrregb}) and (\ref{fragradius}) is also mild; the explicit dependence $1/f$ in Eq.~(\ref{tmergeq}) is still present but  will cancel when   compared to the spin-aligment time (Sec.~\ref{fixedparameters}).
On the other hand, the dependence on $q$ plays a crucial role when comparing each spin-alignment time with the inspiral time-scale, and is therefore the main subject of this study. 

\subsection{BH spin alignment}
\label{barpet}
The circumprimary and cirbumsecondary discs interact with the BHs through the Bardeen-Petterson effect. \cite{bardeen75} showed that a viscous disc initially misaligned with the equatorial plane of a spinning BH naturally relaxes to a coplanar state in the inner regions, while the outer disc may retain its original misalignment. \cite{rees78} realized that, by Newton's third law, the outer disc must react by pulling the BH towards  complete alignment (or antialignment) of the  spin with the orbital angular momentum of the outer disc itself. 

Angular momentum is initially transferred  from the spin to the inner disc trough relativistic Lense-Thirring precession  and finally to the outer disc by the propagation of warps, i.e. vertical shearing by close, misaligned, gas rings \citep{scheuer96,LP06,martin07b}. Warp propagation is ruled by a vertical viscosity coefficient $\nu_2$, which is generally different than the  kinematic viscosity coefficient $\nu$ introduced  above. 
As done for $\alpha$, let us introduce a vertical-viscosity coefficient $\alpha_2$ such that $\nu_2 = \alpha_2 c_s H$ \citep{pappringle83}.
In the small-warp  --which in our case actually means small-misalignment-- limit, the warp-propagation coefficient is related to the kinematic viscosity by \citep{pringle92,ogilvie99} 
\begin{align}
\alpha_2 = \frac{1}{2\alpha} \frac{4(1+7\alpha^2)}{4+ \alpha^2}\;,
\label{alpha2linear}
\end{align}
and, in particular, it is independent of the misalignment $\varphi$ between the inner disc and the outer disc. A full non-linear theory of warp propagation has been computed by \cite{ogilvie99} and later verified numerically by \cite{LP10}. Non-linearities introduce a qualitatively new dependence\footnote{The warp-propagation coefficient actually depends on the radial derivative of the local inclination of the disc $\psi$, see Eq.~(1) in \cite{LP10}. Here we implement the same approximation $\psi\sim\varphi$ as already done by \cite{LG13}.} on $ \varphi$, which can lower the value of $\alpha_2$ by a factor of $\sim 7$  for large misalignment angles (see Fig.~1 in \citealt{LG13}). In this paper we  consider the full non-linear expression ${\alpha_2(\alpha,\varphi)}$ as derived by \cite{ogilvie99},
which reduces to Eq.~(\ref{alpha2linear}) for $\varphi\ll 1$. 

Lense-Thirring precession efficiently aligns the disc up to the Bardeen-Petterson radius $R_{BP}$, defined to be the disc location where the inverse of the Lense-Thirring precession frequency  \citep{wilkins72} 
\begin{align}
\Omega_{LT}(r)= 2 \frac{G^2 M^2 a}{c^3 r^3}
\label{omegalt}
\end{align}
equals the warp propagation time $t_{\nu_2}(r)=r^2/ \nu_2$, i.e 
\begin{align}
R_{\rm BP} = 2^{2/3} \left( \frac{a}{\alpha_2}\right)^{2/3} \left(\frac{H}{r}\right)^{-4/3} \left(\frac{GM}{c^2}\right)\,.
\label{rbp}
\end{align}
For a single BH-disc system, $R_{BP}$ coincides with the maximum warp location (\emph{warp radius}) $R_W$ and marks the boundary between the (quickly aligned) inner disc and the (still aligning) outer disc. The time-scale over which the outer disc  finally aligns the BH spin can be found by computing the torque acting on the disc at  $R_W$ \citep{natarajan98}. A single BH of mass $M$ and dimensionless spin  $a$ aligns with the angular momentum of the surrounding accretion disc within \citep{scheuer96,natarajan98,LG13} 
\begin{align}
t_{\rm{al}}\simeq 3.4 \;  \frac{M}{\dot M}\;\alpha \;\left( \frac{a}{\alpha_2
}   \frac{H}{r} \right)^{2/3}  ,
\label{talign}
\end{align}
where $\dot M$  is the accretion rate of the circumprimary/circumsecondary disc  and ${H}/{r}$ its aspect ratio evaluated  at the warp radius.
We note here that the alignment time $t_{\rm al}$ is sensibly smaller that the growth time $M/\dot M$ for reasonable viscosities $\alpha\sim 0.1$ and aspect ratios $H/r\sim0.001$. BH mass and spin magnitude
 can be therefore considered fixed during the alignment process (\citealt{kingkolb}; see Sec.~\ref{discussion}).

The Bardeen-Petterson effect can  drive the BH spin towards either alignment or antialignment with the outer disc.  \cite{KLOP} showed that the system antialigns if
\begin{align}
\theta > \pi/2 \quad {\rm and}  \quad L(R_W)<2 S\;,
\end{align}
where  $\theta$ is the angle between the BH spin and the angular momentum of the outer disc, $L(R_W)$ is the angular momentum of the inner disc (i.e. inside the warp radius) and $S$ is the BH spin.
The BH spin aligns with the outer disc if any of the two conditions above is not satisfied.
Once $\theta$ is provided (cf. Sec.~\ref{results}),
the misalignment $\varphi$ between the inner-disc angular momentum  and the outer-disc angular momentum is given by $\varphi=\theta$ in the aligned case, while $\varphi=\pi-\theta$ if the system tends towards antialignment. Note that even in cases where the BH spin antialigns with its own disc, the net effect is always to reduce the misalignment with the binary plane \citep{KLOP}.

\subsection{Effect of the companion on  disc-spin alignment}
\label{companion}
\begin{figure*}
\centering
\includegraphics[width=\textwidth]{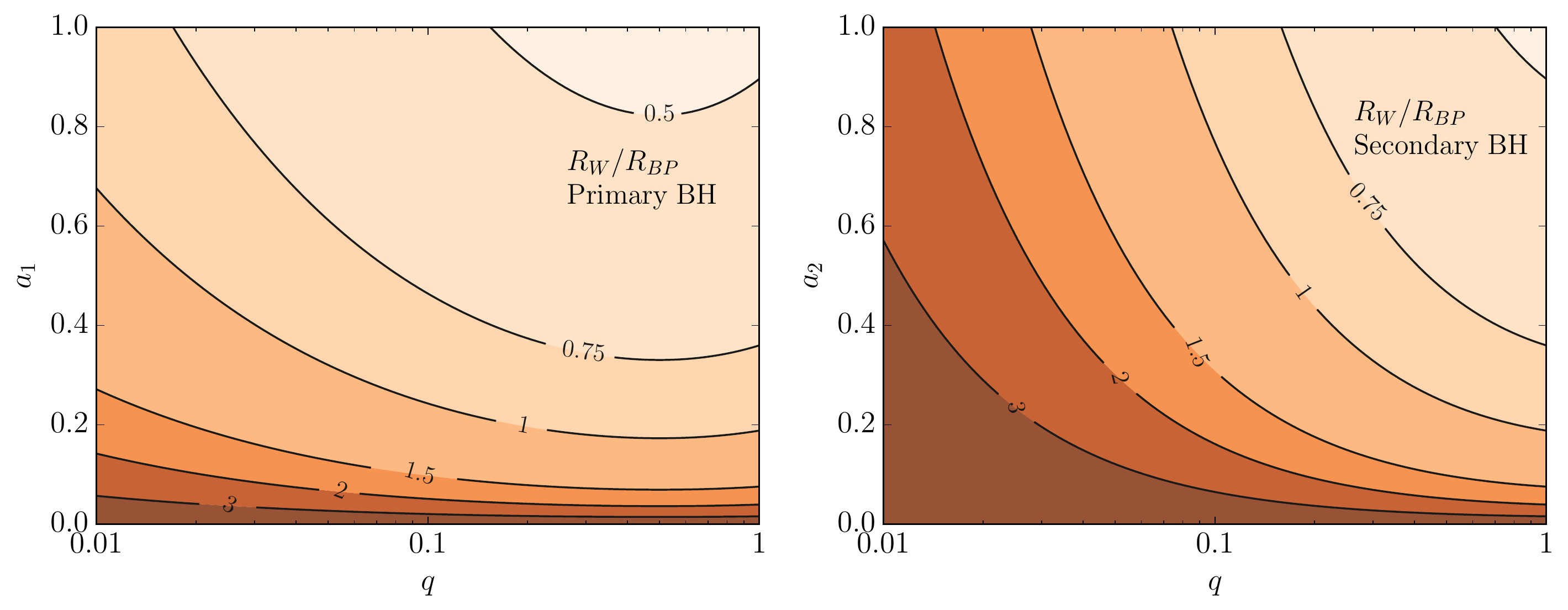}
\caption{Effect of the companion on the location of the warp radius. Primary and secondary BHs are considered on the left-hand and right-hand panel respectively. Contours map the ratio between the maximum-warp location $R_{W}$ (where the external torque from the companion matches the Lense-Thirring torque) and the Bardeen-Petterson radius $R_{BP}$ (where the Lense-Thirring time equals the warp propagation time) derived for an isolated BH.  The dependence on the binary mass ratio $q$ and the spin of the aligning BH (either $a_1$ or $a_2$) is reported on the $x$ and $y$ axes. A companion BH is expected to speed up the alignment process up to a factor of $\sim 2$ in most of the parameter space, meaning a $\mathcal{O}(1)$ uncertainty in the estimate of the alignment time. This figures have been produced taking $M_{\rm bin}=10^7 M_\odot$, $H/r=0.001$, $f=0.1$, $\alpha=1/2\alpha_2=0.2$, $\beta=1$ and $R=R_{\rm f}$ [see Eq.~(\ref{rw_over_rbp})].}\label{comp_contour}
\end{figure*}

So far we have only considered the alignment of a single BH with its surrounding accretion disc. Here we discuss the effect of a far ($R \gg R_{W}$) companion on the alignment process. Such effect --neglected in our previous study \citep{LG13}-- has been recently pointed out by \cite{miller13} in the supermassive BH binary case, while \cite{martin09} have previously considered the same interaction for stellar-mass BHs with stellar companions. For a further study, see \cite{tremaine14}.

If the aligning BH is part of a binary system, the gravitational potential felt by an orbiting gas ring is perturbed by the presence of the companion  \citep{lubow00,ogilvie01}. The binary gravitational potential can be expanded in series of $r/R$ (where $r$ is the distance of the gas ring from the BH 
 and $R$ is the binary separation; see  e.g. \citealt{katz82}):  
to leading order, the resulting torque is perpendicular to the angular momentum of the gas ring $\mathbf{L}$, causing  precession about the angular momentum of the binary $\mathbf{L}_{\rm bin}$. The precessional frequency can  be obtained by averaging the torque over the binary orbital period and reads\footnote{We do not quote the sign of the precession frequency, because it only sets the precession direction about $\mathbf{L}_{\rm bin}$ which is not important for our order-of-magnitude estimate.} \citep{petterson77}
\begin{align}
\Omega_{C}(r) = \frac{3}{4} \frac{G M_{\rm c}}{R^3} \left( \frac{r^3}{GM}\right)^{1/2} \beta\;,
\label{omegac}
\end{align}
where $\beta= |\mathbf{\hat L}_{\rm bin}\cdot \mathbf{\hat L}|$, $M$ is the mass of the aligning BH and $M_{\rm c}$ is the mass of the companion.  Note that in our notation $M_{\rm c} = q M$ in the case of the primary BH, but $M_{\rm c}=M/q$ when the alignment of the secondary is considered.

If a spinning BH is part of a binary system, both Lense-Thirring and companion-induced precession are present. The companion drives the system towards alignment with the angular momentum of the binary, which tracks the plane of the circumbinary disc (see Sec.~\ref{discussion} on this point).  At the same time, the inner disc is being aligned to the BH spin by Lense-Thirring precession.
In practice, the companion reduces the frame-dragging efficiency: material could stay misaligned with the BH spin at closer locations, thus speeding  the alignment process up \citep{miller13}. 
This effect can be quantified by computing the locations  at which the two contributions are equally important. The Lense-Thirring time $\Omega^{-1}_{LT}$ equals the warp propagation time $t_{\nu_2}$ at $R_{BP}$, as given by Eq.~(\ref{rbp}). On the other hand, the disc is now expected to be maximally warped at the warp radius $R_W$, where the Lense-Thirring contribution matches the companion one  $\Omega_{LT} = \Omega_C$ \citep{martin09}. From Eq.~(\ref{omegalt}) and (\ref{omegac}) one gets (\citealt{martin09,miller13})
\begin{align}
R_{W}=  \left(\frac{8 a }{3 \beta} \frac{M}{M_{\rm c}}\right)^{2/9}
R^{2/3} \left(\frac{GM}{c^2} \right)^{1/3} \,.
\end{align}
If $R_{W}\gtrsim R_{BP}$, the companion term can be neglected and the closer location at which misaligned material can be found is still $\sim R_{BP}$.  The alignment speed-up discussed by \cite{miller13} is relevant if $R_{W}\lesssim R_{BP}$, because warped regions are present closer to the hole.  At $R=R_{\rm f}$ (cf. Sec.~\ref{fragmentation}), we find
\begin{align}
\begin{split}
\frac{R_{W}}{R_{BP}}&\simeq 0.48 \:\: \beta^{-2/9}
a^{-4/9}
    \left[ \frac{M+M_{\rm c}}{2 \;M_{\rm c}^{1/3} M^{2/3}}\right]^{2/3} 
\left( \frac{M+M_{\rm c}}{10^7 M_\odot} \right)^{-52/81} \\
&\times
\left(\frac{H/r}{0.001} \right)^{4/3}
\left( \frac{f}{0.1}\right)^{-16/81}
\left( \frac{\alpha}{0.2}\right)^{-26/81}
\left( \frac{\alpha_2}{1/2\alpha}\right)^{2/3} \,.
\end{split}
\label{rw_over_rbp}
\end{align}
Fig.~\ref{comp_contour} shows the dependences of  ${R_{W}}/{R_{BP}}$ on the binary mass ratio and the spin magnitude of the aligning BH, both for primaries and secondaries. Slowly rotating BHs are less affected by the presence of  a companion because the spin set the magnitude of the frame-dragging term.  For fixed total mass $M_{\rm bin}=M+M_{\rm c}$, primaries are more sensible to the companion than secondaries, because their gravitational radius is larger and Lense-Thirring precession can be matched more easily by the additional precession term.

In this paper we use the simple expression reported in   Eq.~(\ref{talign}) to compute the spin-alignment time, as formally obtained for an isolated BH-disc system. 
Our analysis [Eq.~(\ref{rw_over_rbp}) and Fig.~\ref{comp_contour}] shows that  the position of the warp radius can be modified by a factor of $\sim 2$ if the BH is part of a binary system. The alignment time $t_{\rm al}\propto R_W^{11/10}$ \citep{natarajan98,miller13} can therefore only be lowered by a factor of $\sim$few. 
From Eq.~(\ref{rw_over_rbp}), this assumption may not be valid if (i) the binary is very massive  $M_{\rm bin} \gtrsim 10^7 M_\odot$, (ii) the individual discs are  thinner than  the circumbinary disc at the fragmentation radius $H/r \lesssim  0.001$, (iii) the binary accretion rate is close to the Eddington limit $f>0.1$. 
 A more complete understating of the alignment process in BH binary systems requires explicit integrations of the angular momentum equation \citep{martin09}. This goes beyond the scope of this work, which instead focuses on getting an estimate for the alignment time-scale.

\subsection{Cavity pile-up and differential accretion}
\label{diffacc}
\begin{figure*}
\centering
\includegraphics[width=\textwidth]{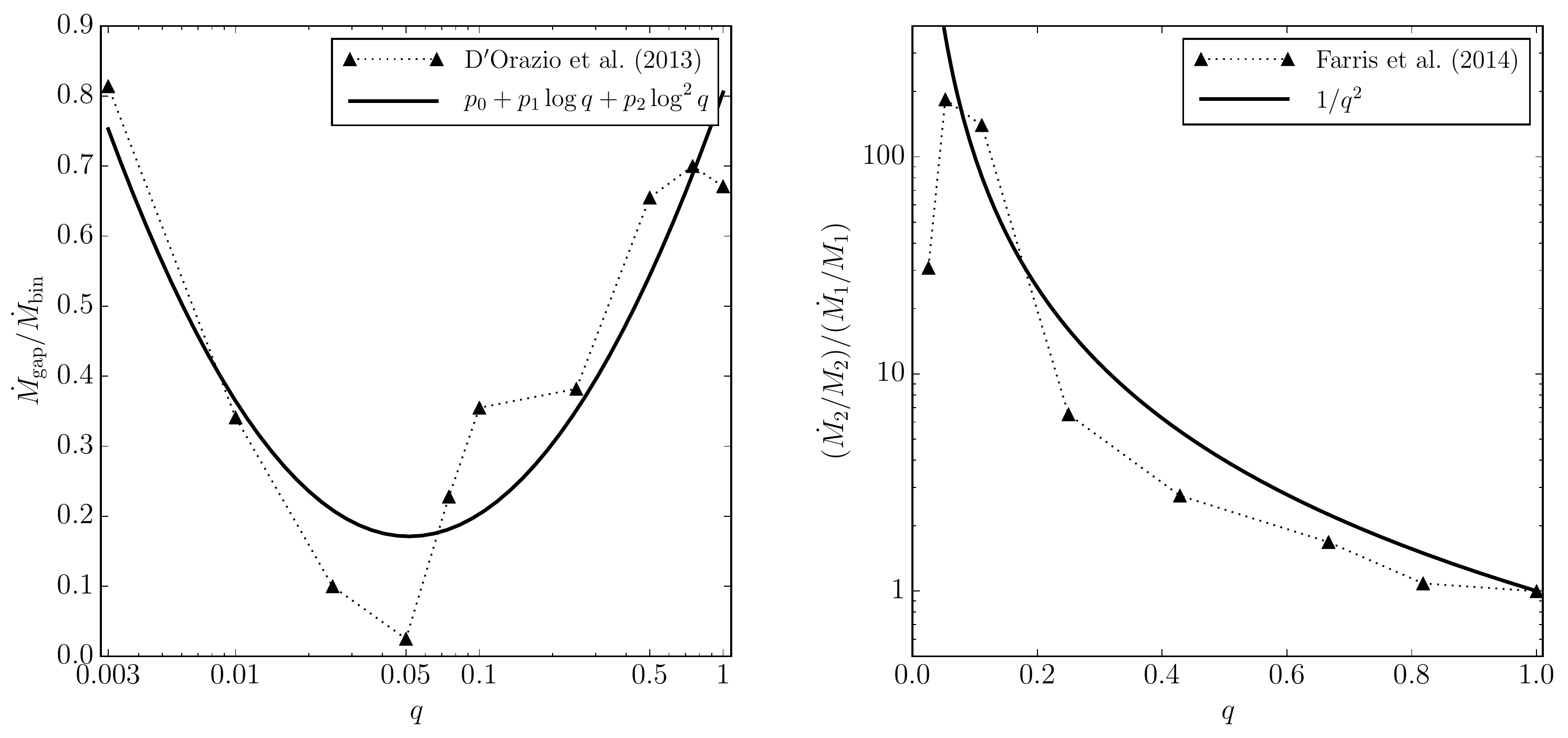}
\caption{Numerical fits to hydrodynamical simulations to compute the accretion rates of the two members of a BH binary. Left-hand panel shows the fraction of the accretion rate $\dot M_{\rm gap} / \dot M_{\rm bin}$ that penetrates trough the disc cavity and reaches one of the two binary members. A quadratic interpolation is performed to numerical 
results
by \protect\cite{dorazio13} here reported in Eq.~(\ref{dorazio_interp}). On the right-hand panel, the binary accretion rates is broken down between the primary and the secondary BHs. The ratio of the accretion times  in the systems simulated by \protect\cite{farris14} appears to be described by the simple prescription $( \dot M_2/M_2)/( \dot M_1 / M_1)=1/q^2$ in  Eq.~(\ref{farrissimple}).}
\label{twofit}
\end{figure*}

The accretions rates of the individual discs $\dot M_1$  and $\dot M_2$ depend on the circumbinary-disc accretion rate  $\dot M_{\rm bin}$, since the formers are fed by the latter. Here we develop a simple prescription to link these three quantities.

Accretion from the outermost regions of the circumbinary disc on to the binary BH is suppressed because of either the perturbation of the secondary (low-$q$ regime) or the two-body central potential (high-$q$ regime). Therefore, the binary may only accrete at a lower accretion rate $\dot M_1+ \dot M_2 \leq \dot M_{\rm bin}$. Mass tends to pile up at the outer edge of the cavity created by the binary itself: accretion --and therefore spin alignment-- is still possible if  gas streams can penetrate the cavity and reach the BHs.   We call $\dot M_{\rm gap}$ the mass accretion rate that overcomes the cavity pile-up: this gas will sooner or later accrete on to either the primary or the secondary BH, i.e.
\begin{align}
\dot M_{\rm gap} = \dot M_1 + \dot M_2\;.
\label{mdotsum}
\end{align}

Gas-stream propagation is an intrinsically multi-dimensional non-linear phenomenon that requires dedicated hydrodynamical simulations to be studied in detail. In particular, the dynamics of gas accretion through the cavity is strongly dependent on the binary mass ratio $q$, since qualitatively different regimes are present. \cite{macfadyen08} first discussed equal-mass binary simulations, while the $q=1/2$ case has been presented by \cite{hayasaki07} and extended to $q=1/3$ by \cite{cuadra09} and \cite{roedig12}.  
  A recent major improvement has been made by \cite{dorazio13} and \cite{farris14} who extensively studied the dependence on $q$ of the mass rate overcoming the cavity pile-up.
  
\cite{dorazio13} present 2D hydrodynamical simulations in the range $0.003\leq q \leq 1$ assuming fiducial values $\alpha=0.01$ and $H/r=0.1$. Accretion on to the binary is indeed limited to narrow gas streams and it is typically suppressed by a factor of 2-5 when compared to a single-BH disc of the same mass. They detect the presence of two physical regimes:
\begin{enumerate}
\item For high mass ratios $0.05\lesssim q\leq1$, the presence of the binary strongly modulates the streams. Streams are generated by deviations from spherical symmetry in the binary potential: more asymmetry is present for equal mass binaries that therefore show less mass pile-up at the cavity edge and more binary accretion. In such regime, the ratio  $\dot M_{\rm gap}/ \dot M_{bin}$ is expected to increase with $q$.
\item In the low-mass ratio regime $q \lesssim 0.05$, the secondary BH quickly swipes through the disc accreting most of the material coming from large distances: a single gas  stream is present feeding the secondary BH. Such effect gets more pronounced when the mass ratio is lower and consequently the ratio  $\dot M_{\rm gap}/ \dot M_{\rm bin}$ decreases with $q$.
\end{enumerate}
Such results are shown in the left-hand panel of Fig.~\ref{twofit}, were the ``Mid$\Delta r$-Lo$\Delta\phi$'' simulations by \cite{dorazio13}  are considered. The minimum in $\dot M_{\rm gap}/ \dot M_{\rm bin}$ separates the two physical regimes just described above. 
We interpolate the results from the simulations performed by \cite{dorazio13} with the ansatz
\begin{align}
\frac{\dot M_{\rm gap}}{\dot M_{\rm bin}} = p_0 + p_1 \log(q) + p_2 \log^2(q)
\label{dorazio_interp}
\end{align}
and best-fitting coefficients $p_0=0.8054$, $p_1=0.9840$ and $p_2= 0.3818$. For a given circumbinary-disc accretion rate $\dot M_{\rm bin}$, Eq.~(\ref{dorazio_interp}) specifies the mass rate $\dot M_{\rm gap}$ which overcomes the cavity pile-up and accretes on to \emph{either} the primary \emph{or} the secondary BH. 
The subsequent study by \cite{farris14} found that the ratio between the total accretion rate on to either one of the two BHs and  the accretion rate  on to a single BH of the same total mass may exceed unity, thus casting doubts on whether such fraction can be interpreted as ${\dot M}_{\rm gap}/{\dot M}_{\rm bin}$. To bracket this uncertainty, we use Eq.~(\ref{dorazio_interp}) as our reference model but we also study an additional variation where we fix ${\dot M}_{\rm gap}={\dot M}_{\rm bin}$ (cf. Sec.~\ref{fixedparameters}).

\cite{farris14} recently performed 2D grid simulations (assuming $H/R=0.1$ and $\alpha=0.1$), specifically addressing the feeding of the individual discs from  streams penetrating the cavity. They systematically find that the secondary BH accretes faster than the primary, mainly because the former orbits closer to the cavity edge. Their results are here reported in the right-hand panel of Fig.~\ref{twofit}, where  the ratio of the accretion times $\dot{M_i}/M_i$  ($i=1,2$) is showed as a function of the binary mass ratio $q$. Symmetry implies ${\dot M_1} \sim {\dot M_2}$ for binaries with high mass ratios, while lower values of $q$ show pronounced differential accretion in favour of the secondary. A qualitatively different regime is detected for the lowest of their simulated cases $q=0.025$: the cavity is not efficiently cleared by the secondary BH, and mass from the circumbinary disc directly flows inwards forming a large circumprimary disc. As pointed out in Sec.~\ref{gasdriventime}, such change in the dynamics of the system is expected for lower mass ratio, where the disc should form a small annular gap rather than a large hollow cavity. 
To directly reach the circumprimary disc, gas should be able to flow past the secondary escaping its gravitational attraction.  As pointed out by \cite{farris14} themselves, the actual turning point in $q$ is  likely to be highly dependent on the thickness of the disc and possibly on the viscosity.
As shown recently by \citet{Young2015} in the context of binary stars, direct flowing from the circumbinary to the circumprimary disc is easier for thicker discs, where the stronger pressure forces can make part of the material ``skirt'' the Roche lobe of the secondary, eventually reaching the primary Roche lobe and being captured by its gravitational attraction. Due to such uncertainties, in this work we deliberately ignore the onset of such low-$q$ regime when considering differential accretion. 
The growth-time ratio 
presented by \cite{farris14} appear to be well approximated by  (see Fig.~\ref{twofit}, right-hand panel)
\begin{align}
\frac{\dot M_2/ M_2}{\dot M_1/  M_1} = \frac{1}{q^2}\;.
\label{farrissimple}
\end{align}
Due to such pronounced differential accretion, the prescription here presented may formally predict super-Eddington rates for the secondary BHs in the low-$q$ regime. This has no relevant impact on our model:  defined the secondary Eddington ratio to be $f_2=\epsilon t_{\rm Edd} {\dot M_2} / M_2 = (1+q) {\dot M_2}  \epsilon f / q  {\dot M_{\rm bin}} $,  the Eddington limit $f_2=1$ is only marginally reached for very high circumbinary-disc accretion rates ($f \sim 1$) and low mass ratio $q \lesssim 0.05$ (assuming a typical accretion efficiency $\epsilon\sim0.1$).  Lower values of $f$ shift the critical mass
ratio at which the Eddington limit is formally reached to even lower
values.
Finally, we note that, as in Sec.~\ref{gasdriventime}, the thickness values explored by the simulations considered in this section are significantly higher than those expected for massive-BH binaries (Sec.~\ref{fragmentation}).

\begin{figure}
\centering
\includegraphics[width=0.45\textwidth]{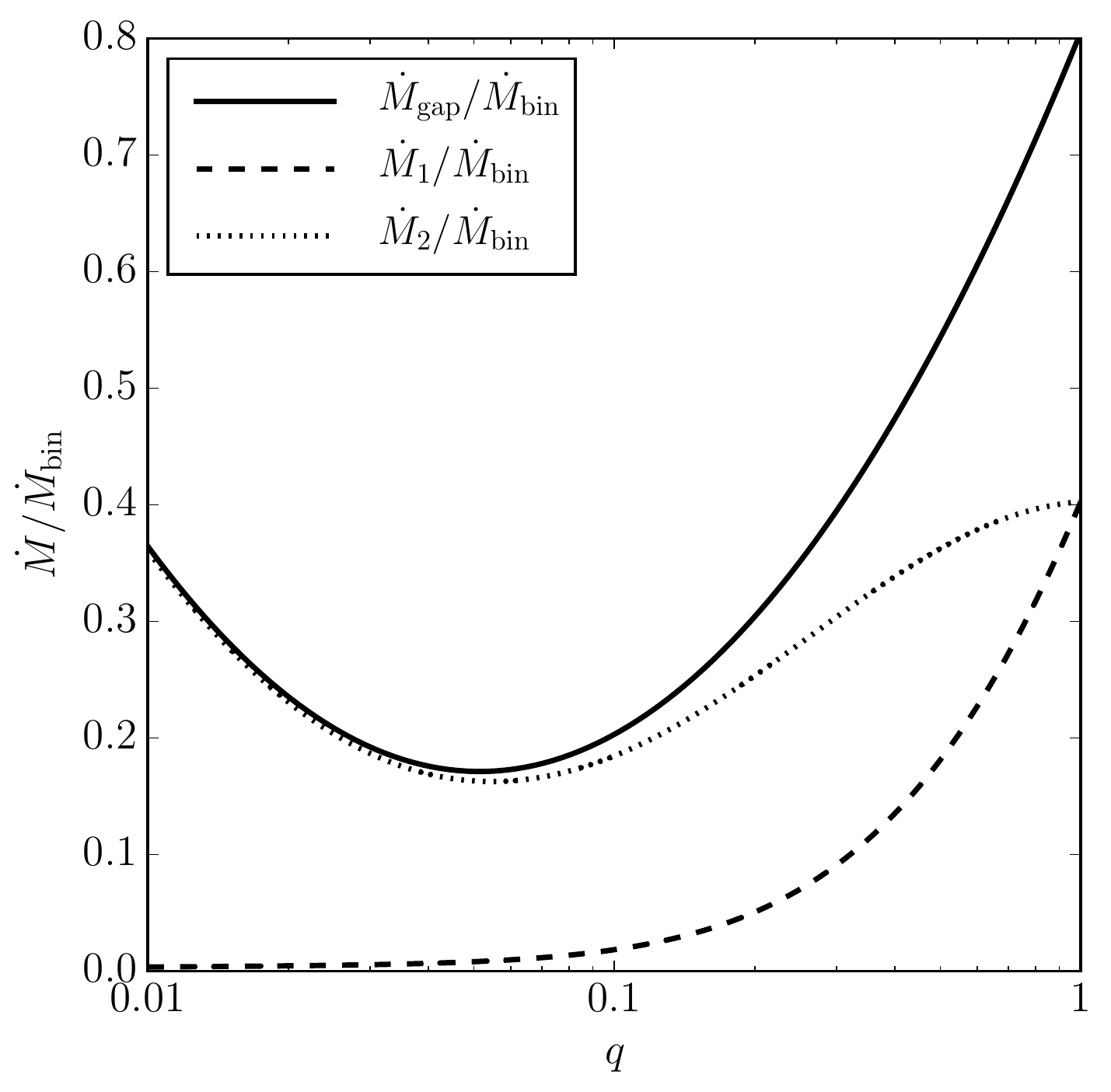} 
\caption{Combined effect of mass pile-up at the edge of the disc cavity and differential accretion in unequal-mass binaries. Prescriptions for the accretion rates presented in  Eqs.~(\ref{dorazio_interp}-\ref{dotm1dotm2}) and Fig.~(\ref{twofit}) are here summarized. As a function of $q$, we show the fraction of the accretion rate of the outer circumbinary disc $\dot M_{\rm bin}$ that   (i) overcomes the cavity-pile up and accretes on to the binary (${\dot M_{\rm gap}}$, solid); (ii) is captured by the secondary BH which is clearing the cavity (${\dot M_2}$, dashed); (iii) finally accretes on to the primary BH  (${\dot M_1}$, dotted).
}
\label{diffaccfig}
\end{figure}
From Eq.~(\ref{mdotsum}) and (\ref{farrissimple}) one gets
\begin{align}
\dot  M_{1}=\frac{q}{1+q}\dot{M}_{\rm{gap}}\;, \qquad
\dot M_{2}=\frac{1}{1+q}\dot{M}_{\rm{gap}}\;.
\label{dotm1dotm2}
\end{align}
Fig.~(\ref{diffaccfig}) combines the results presented in Eq.~(\ref{dorazio_interp}) and Eq.~(\ref{dotm1dotm2}). For equal-mass binaries, $\sim 80\%$ of the incoming mass may accrete on to the binary and it is equally distributed between the two binary members. Unequal-mass binaries present differential accretion $\dot M_1<\dot M_2$ that grows stronger as  $q$ is decreased. When $q$ is increased from  $q\sim 0.2$ to unity, gas streams start to flows towards the primary and $\dot M_2/ \dot M_{\rm bin}$ consequently flattens. On the other hand, if $q$ is decreased from $q\sim 0.05$ to $0.001$, the secondary orbit gets closer to the inner edge of the cavity \citep{dorazio13}: more mass can overcome the cavity pile-up and it is almost entirely accreted by the secondary. 

\section{Results: differential misalignment}
\label{results}
In this section we compare the spin-alignment time and the inspiral time.  
We first outline the regions of the parameter space where misalignments are foreseen (Sec.~\ref{fixedparameters}); secondly, we  fold our model into synthetic supermassive-BH binary populations (Sec.~\ref{cosmological}); and we finally  present a preliminary study to address the impact of our findings on the occurrence of large post-merger kicks (Sec.~\ref{kicks}).

\subsection{Misaligned primary BHs} 
\label{fixedparameters}

\subsubsection{Fiducial values of the parameters}

The circumbinary disc properties enter the inspiral time $t_{\rm in}$ while the primary/secondary alignment times $t_{\rm al}$ are set by individual-disc parameters. The ratio $t_{\rm al}/t_{\rm in}$ in general depends on the binary separation $R$, the three disc aspect ratios $H/r$, the gas viscosity $\alpha$, the accretion rates of the circumbinary $\dot M_{\rm bin}$  and the individual discs $\dot M_{1,2}$, the BH masses $M_1$ and $M_2$ (or equivalently $q$ and $M_{\rm bin}$), the orientation angles $\theta_1$ and $\theta_2$, and the BH spin magnitudes $a_1$ and $a_2$. We first specify a fiducial model by taking likely values of all these parameters and we later perform a small parameter study around such model. Table \ref{param_all} summarizes the values we assume for the parameters, highlighting the differences with the next section. We discuss our choices  as follows.

\begin{table}
\center
\begin{tabular}{c@{\hskip 0.2in}|@{\hskip 0.2in}c @{\hskip 0.2in}c}
Parameter & Fiducial model & Synthetic distributions  \\ \hline \hline
$q$ & Free parameter&Power-law distributions\\
$M_{\rm bin}$ & Not relevant & Not relevant\\
$a_1$, $a_2$&1&Either 1 (E) or 0.1 (C)\\
$\theta_1$, $\theta_2$ & Extremize over & Random variables\\
$R$ &  $R_{\rm f}$ (fragmentation)  & $R_{\rm f}$ (fragmentation) \\
$H/r$ & $0.001$ & $0.001$\\
$\alpha$ & $0.2$ & $0.2$\\
$f$ & Not relevant & Not relevant\\
$\dot M_1$, $\dot M_2$& Eqs.~(\ref{dorazio_interp}) and (\ref{dotm1dotm2}) &  Eqs.~(\ref{dorazio_interp}) and (\ref{dotm1dotm2}) \\
\end{tabular}
 \label{param_all}

\caption{Choice of the binary and disc parameters in our time-scale comparison for both the fiducial case (Sec.~\ref{fixedparameters}) and the cosmologically motivated distributions (Sec.~\ref{cosmological}). }
\end{table}

\begin{itemize}
\item
As detailed in Sec.~\ref{fragmentation}, a rather conservative assumption can be made by evaluating the inspiral time at the fragmentation radius $R_{\rm f}$. This is a measurement of  largest separation where the inspiral can be driven by interaction with a gaseous environment and is typically believed to be the bottleneck of the whole binary evolution.
\item
In our fiducial model we fix the aspect ratios of all discs to $H/r=0.001$. As reported in Eq.~(\ref{hrregb}) for the circumbinary disc, the aspect-ratio dependences on the other parameters (namely the viscosity, the accreting mass and the accretion rate) are not crucial to evaluate the inspiral time-scale, and will be here neglected for simplicity. For the same reason, we assume the individual discs to share the same aspect ratio of the circumbinary disc (cf. the analogous assumption made by \citealt{miller13}).

\item
Unless specified, we fix $\alpha=0.2$. A parametric study on the viscosity has already been presented previously \citep{LG13} and the alignment process has been found to be overall quite independent of $\alpha$.   Negative azimuthal viscosities are formally predicted by the non-linear warp propagation theory for $\alpha \lesssim 0.1$ and large misalignments $\varphi$ \citep{ogilvie99,LG13}: the evolution of the disc in these cases is unclear and out of the scope of this study (see \citealt{nixonking12} and \citealt{nixon13} for extensive discussions). 
\item
As described in
 Sec.~\ref{diffacc}, the BH accretion rates  $\dot M_1$ and $\dot M_2$ are related to the circumbinary disc accretion rate $\dot M_{\rm bin}$, conveniently expressed through the dimensionless quantity $f$ in Eq.~(\ref{deff}).  In our fiducial model we implement Eqs.~(\ref{dorazio_interp}) and (\ref{dotm1dotm2}).  Once $H/r$ is fixed, 
the alignment likelihood  $t_{\rm al}/t_{\rm in}$ is independent of $f$ because both times scale as $1/f$ [cf. Eqs.~(\ref{tmergeq}) and (\ref{talign})]. This is a point of improvement over our previous estimate \citep{LG13}, where an effective dependence on $f$ was introduced when decoupling the inspiral and the alignment processes. For concreteness, the overall scale of Fig.~\ref{band} below is computed assuming $f=0.1$.

\item
Within our assumptions, both the inspiral and the alignment times are independent of the binary total mass $M_{\rm bin}$. This is compatible with  \cite{haiman09} when $t_{\rm in}$ is evaluated at the fragmentation radius. 

\item
The orientations angles $\theta_1$ and $\theta_2$ set the warp efficiency $\alpha_2$ (Sec.~\ref{barpet}) and their dependence is the main point raised by \cite{LG13}. 
In the following, we bracket such uncertainties extremizing $t_{\rm al}$ over all possible orientations.

\item
For simplicity, we consider maximally spinning BHs (\mbox{$a_1=a_2=1$}) unless specified.  The status of supermassive-BH spin measurements has been recently reviewed by \cite{reynolds13}: some highly spinning BHs are found, but the current statistic is too low to provide a complete picture of the spin magnitude distributions. The effect of the spin magnitude on the alignment likelihood can however be easily predicted, because the alignment time scales as $t_{\rm al}\propto a^{2/3}$, cf. Eq.~(\ref{talign}).
\end{itemize}

\subsubsection{Predicted time-scales}
The key dependence of the problem is the one on the binary mass ratio $q$, which both marks the onset of different inspiral regimes and sets the importance of differential accretion. 
\begin{figure}
\includegraphics[width=0.98\columnwidth]{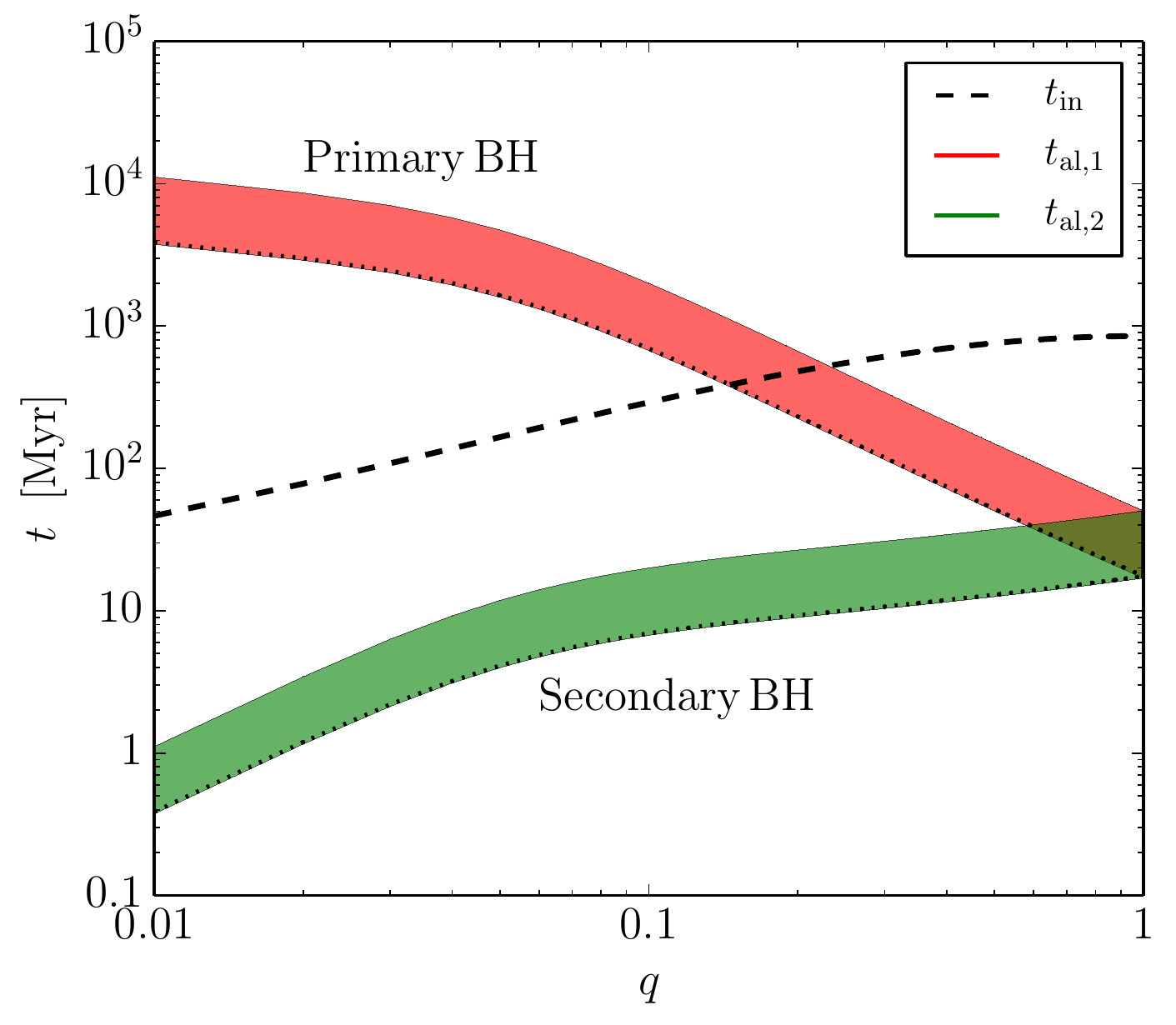}
\caption{Comparison between the alignment times and the inspiral time as a function of the binary mass ratio $q$. Gas interactions have time to align the BH spins with the orbital angular momentum of the binary if the alignment time $t_{\rm al}$ (shaded areas) is smaller than inspiral time $t_{\rm in}$ (dashed line). While secondary BHs (lower, green area)  aligns for all mass ratios, this is not the case for the primary members (upper, red area) which retain their initial misalignments if $q$ is low enough. Our fiducial model is assumed here: $H/r=0.001$, $\alpha=0.2$, $f=0.1$, and maximally spinning BHs $a_1=a_2=1$.  Warp non-linear propagation theory introduces uncertainties (thus the shaded areas) of a factor of $\sim~2.5$ in the alignment times. The corresponding times obtained with the linear theory are shown with dotted lines for comparison and always underestimate the non-linear result.}
\label{band}
\end{figure}
Fig.~\ref{band} shows the inspiral and the alignments times as a function of $q$ for our fiducial set of parameters. The uncertainty in the initial misalignments $\theta_i$ causes the alignment times to appear as stripes in the figure, rather than lines. For comparison, we also show (dotted lines) the behaviour predicted by the linear warp-propagation of Eq.~(\ref{alpha2linear}) where $t_{\rm al}$ is independent of $\theta_i$. The linear theory underestimates the alignment time by up to a factor of $\sim 2.5$, as already pointed out by \cite{LG13}.
Fig.~\ref{band} illustrates the main result of this paper: while secondaries are found aligned ($t_{\rm al,2}\ll t_{\rm in}$) for every value of $q$,  primary BHs only align if $q\gtrsim 0.2$. Light secondaries may prevent primaries from aligning. If such BHs were misaligned before the disc interactions, these misalignments are carried over to the next stages of the binary evolution. As briefly explored in Sec.~\ref{kicks} this \emph{differential alignment} between the two binary members will affect the subsequent GW-driven inspiral, the merger phase and the properties of the remnant BHs opening for the possibilities of large kicks. 

\begin{table}
\center
\begin{tabular}{c@{\hskip 0.2in}|@{\hskip 0.2in}ccc@{\hskip 0.2in}cc}
Variation & \multicolumn{2}{c}{Primary BH} && \multicolumn{2}{c}{Secondary BH} \\
\hline\hline
Fiducial & [$0.14$ - $0.23$]  &($0.14$)&&[n-n]  &(n) \\
$\alpha=0.3$ & [$0.17$ - $0.28$]  &($0.18$)&&[n-n]  &(n)\\
$\alpha=0.4$ & [$0.19$ - $0.32$]  &($0.21$) && [n-n]  &(n) \\
$\alpha=0.5$ & [$0.21$ - $0.35$]  &($0.23$)&& [n-n]  &(n) \\
$H/r=10^{-4}$ & [$0.07$ - $0.12$]  &($0.07$) && [n-n] & (n) \\
$H/r=10^{-2}$ & [$0.28$ - $0.48$]  &($0.29$) && [n-n]  &(n) \\
$H/r=10^{-1}$ & [$0.60$ - $1$]  &($0.61$) && [$0.02$ - n] & (n) \\
$a_i=0.2$ & [$0.09$ - $0.14$]  &($0.09$) & &[n-n] & (n) \\
$a_i=0.5$ & [$0.12$ - $0.19$]  &($0.12$) & &[n-n] & (n) \\
$a_i=0.8$ & [$0.14$ - $0.22$]  &($0.14$) && [n-n] & (n) \\
$\dot{M}_{\rm{bin}}=\dot{M}_{\rm{gap}}$ & [$0.07$ - $0.12$]  &($0.07$) && [n-n] & (n) \\
\end{tabular}

\caption{Binary mass ratios marking the transition between aligned and misaligned spins. For any variation from our fiducial model [$a_i=1$, $\alpha=0.2$, $H/r=10^{-3}$, $\dot{M}_{\rm gap}$ given by Eq.~(\ref{dorazio_interp})], we report  values $\bar q$ such that BH spins in binaries with $q<\bar q$ are expected to be left misaligned (i.e. $t_{\rm al} > t_{\rm in}$) by gaseous interactions. Values  in square brackets refer to the lower and upper limit of $\bar q$ due to the initial-misalignment uncertainty foreseen using non-linear warp propagation. Values in round brackets show the analogous result when the linear theory is considered, and notably underestimates the value of $\bar q$. Misaligned secondaries are typically not present (as indicated with ``n'') unless some of the parameters are cranked up to unrealistic values.}
\label{tableparam}

\end{table}

A short parametric study around our fiducial model is shown in Table~\ref{tableparam}, where we compute the values of $q$ which mark the onset of the misaligned regime (i.e. where $t_{\rm al}=t_{\rm in}$). As expected \citep{LG13}, the alignment process is rather independent on $\alpha$ with thresholds varying from $q \sim 0.17$ to $0.35$ if $\alpha$ is increased from $0.2$  to $0.5$. Notably, the alignment likelihood is also rather independent on the spin magnitudes $a_1$ and $a_2$, because of the mild scaling of $t_{\rm al}$ [cf. Eq.~(\ref{talign})]. Alignments times are longer for maximally spinning BHs $a_1=a_2=1$ chosen for our fiducial model, but misaligned primaries are predicted for mass ratios $q\sim 0.15$ even when moderately spinning BHs are considered. Perhaps more surprisingly, the alignment process appear to be  strongly dependent on the discs aspect ratio $H/r$ which enters linearly in $t_{\rm in}$ and with a lower power in $t_{\rm al}$. Only primaries with $q>0.6$ have enough time to align their spins in thicker discs $H/r\sim 0.1$, even when maximally rotating BHs are considered. Moreover, if $H/r$ is large enough, the inspiral time may become comparable with the secondary alignment time within the physical uncertainty due to initial spin orientation. As already pointed out (Sec.~\ref{gasdriventime}, se also Sec.~\ref{discussion} below), the disc thickness is one of the main uncertainties in the current modelling of binary-disc interactions. Details of the gas streams leaking through the disc cavity have also a notable effect: the largest value of $q$ where misalignment is foreseen drops down to $\sim 0.12$ if all of the gas of the circumbinary disc ends up being accreted by either one of the two BHs (i.e. if $\dot{M}_{\rm{bin}}=\dot{M}_{\rm{gap}}$, cf. \citealt{farris14}).

\subsection{Cosmogically-motivated distributions}
\label{cosmological}

Our findings are relevant if supermassive-BH binaries with spins and  mass ratios in the misaligned regime are present in nature and detectable. 
While electromagnetic observations already constrained almost a hundred supermassive BH masses \citep{mcconnell13} and a handful BH spins  \citep{reynolds13}, the measurements of the global properties of the supermassive-BH binary population is the main goal of future space-based GW observatories.  eLISA   \citep{eLISA} will detect hundreds binaries per year up to redshift $z \leq 10$ with sufficient signal-to-noise ratio $\mathcal{O} (10-100)$ to measure accurately both individual-source parameters and their statistical distribution \citep{BHeLISA}.

Here we present a simplified analysis to address whether the misaligned-spin regime highlighted above is relevant in such context. 
Publicly available\footnote{{http://www2.iap.fr/users/volonter/LISA\_catalogs.html}} synthetic distributions of merging BH binaries have been developed by the LISA collaboration in the context of the LISA Parameter Estimation Taskforce \citep{PEtask} and later updated by \cite{SGBV}. The authors developed four merger-tree models  of BH evolution, varying over only two ingredients, considered to be the main sources of uncertainty.
\begin{figure}
\includegraphics[width=\columnwidth]{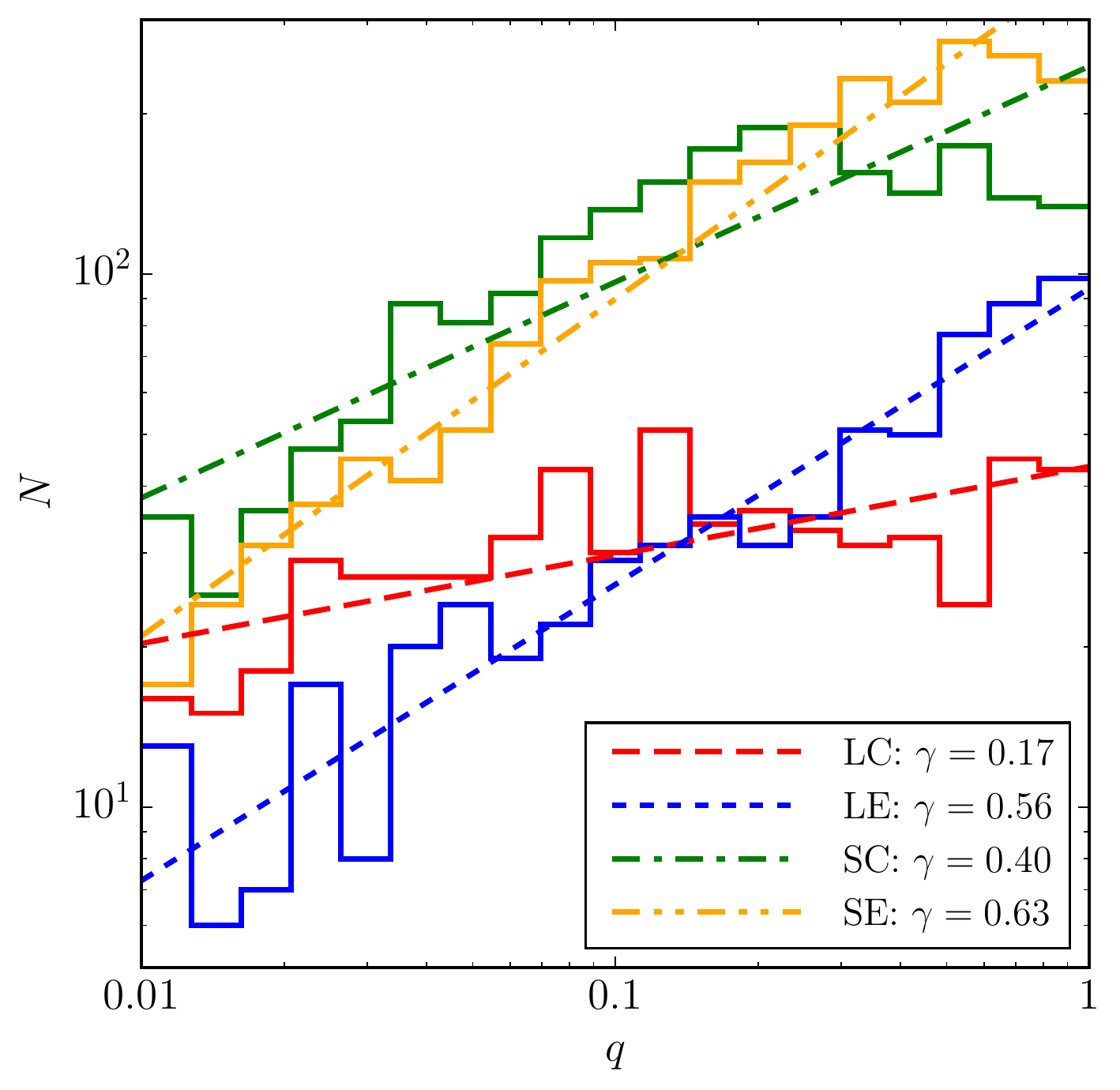}
\caption{Mass ratio distributions in the synthetic supermassive-BH binary populations developed by \protect\cite{PEtask}. Four models are available, for different prescriptions of the accretion geometry ({\bf E}fficient versus {\bf C}haotic) and the BH seeds ({\bf L}arge versus {\bf S}mall). Data sets are binned within the range \mbox{$q\in [0.01,1]$} and fitted with power laws $N \propto q^\gamma$. The best-fitting spectral indexes $\gamma$ are reported in the legend for each model. The histogram normalization has been inherited from the original models and is irrelevant to our purposes.
}
\label{qfit}
\end{figure}
\begin{enumerate}
\item {\it The mass of the BH seeds}. In the small seed scenario ({\bf S}),  first BHs of mass $\sim 100 M_\odot$ are initialized as remnants of Population III stars at $z\sim 20$ and evolved according to \cite{volonteri03}. In the large seed scenario ({\bf L}), BH with mass $10^5 M_\odot$ are formed from gaseous protogalactic discs at  $z\sim 15$ to $\sim 10$ as developed by \cite{begelman06} (see also \citealt{LN06})
\item {\it The accretion geometry}. If accretion efficiently ({\bf E}) occurs on few long episodes, the BHs will generally be spun up during their cosmic evolution \citep{thorne74}. On the other hand, accretion may also happen to be chaotic ({\bf C}), on many short episodes \citep{king06}. In this case, lumps of material accreted in random directions spin on average the holes down.
\end{enumerate}
This approach results into four models, referred to as SE, SC, LE and LC.
Fig.~\ref{qfit} shows the extracted mass ratio distributions, together with power-law fits $N \propto q^\gamma$ in the range $q\in[0.01,1]$. 
We obtain $\gamma=0.17$ for LC, $\gamma=0.56$ for LE, $\gamma=0.40$ for SC and $\gamma=0.63$ for SE.
The spin-magnitude distributions presented by \cite{PEtask} are strongly peaked towards slowly spinning BHs for the C models and $a\sim1$ for the E models.  This is a direct consequence of their simplified accretion treatment, which is either completely coherent or completely chaotic; broader distributions are predicted for more realistic evolutionary models where such assumption is relaxed \citep{barausse12,dotti13,sesana14}.  Spin orientations are not tracked during the cosmic evolution by \cite{PEtask}: spins are assumed to efficiently align in models E, while their directions are kept isotropic in models C. 

\begin{figure}
\begin{center}
\includegraphics[width=\columnwidth]{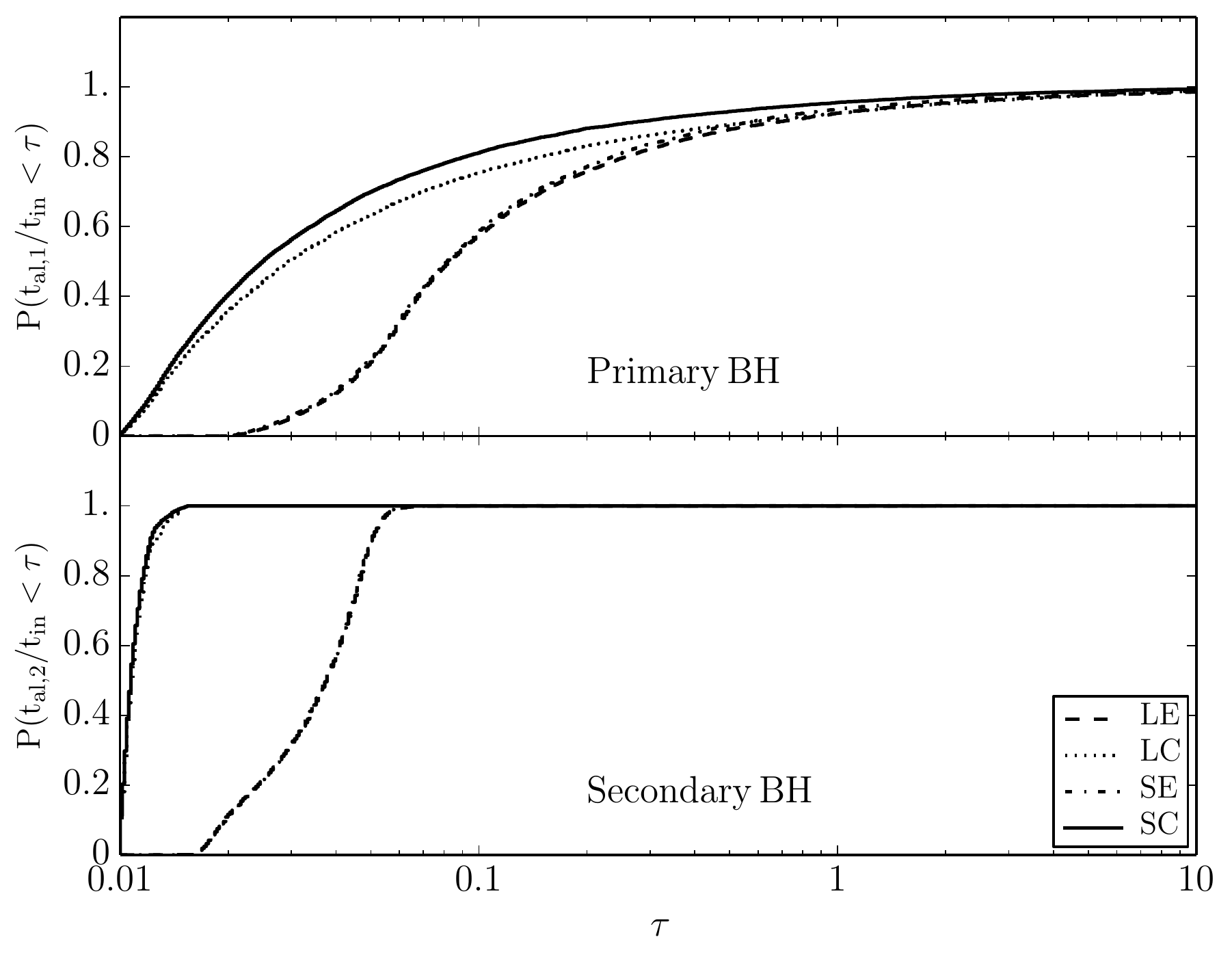}
\caption{Fraction of BH spins in binary systems that align within a factor $\tau$ of the inspiral time $t_{\rm in}$, as predicted using the publicly available distributions by \protect\cite{PEtask}. Alignment predictions using the model presented in this paper can be read at $\tau=1$: all four distributions show that $\sim 8\%$ of BH primaries may fail to align during the gas-driven inspiral, while strong differential accretion quickly aligns all secondaries. Fractions $P$ at larger and lower values of $\tau$ predict the alignment likelihood in case of systematic modelling errors on either the inspiral or the alignment time.}
\label{cumdist}
\end{center}
\end{figure}

Fig.~\ref{cumdist} show the cumulative fraction of aligned BH using these four synthetic BH-binary populations\footnote{We are aware of the inconsistency of our procedure, being the binary mass ratio distributions used here coupled to the spin orientations: at each merger tree level, the properties of the daughter BHs do depend on the spin orientations of their progenitors [cf. Sec.~2.1 in \cite{gerosa15} and references therein].}. We sample the mass ratio $q$ over the fitted power-law distributions from Fig.~\ref{qfit}; spin magnitudes are set to $a_1=a_2=0.1$ in the $C$ models and $a_1=a_2=1$ in the $E$ models, to mimick the strongly peaked distributions of \cite{PEtask}. For simplicity, we fix the disc properties to our fiducial values 
 [$H/r=0.001$, $\alpha=0.2$, $\dot{M}_{\rm gap}$ given by Eq.~(\ref{dorazio_interp})]
and we sample over a uniform distribution in $\cos\theta_i$ to extract values of the alignment time within the initial-orientation uncertainty presented in Sec.~\ref{fixedparameters}.
Fig.~\ref{cumdist} shows, for each value of $\tau$, the fraction of binaries P for which $t_{\rm al}< t_{\rm in} \tau$. Sections at $\tau=1$ correspond to our current model: while all secondaries  align during the inspiral, up to  $\sim 8\%$
of the primaries may not have time to align their spin before merger. This statement appear to be rather independent of the population synthesis model chosen. In particular we find $P(t_{\rm al,1}< t_{\rm in})=0.93$ for LC, $P(t_{\rm al,1}< t_{\rm in})=0.96$ for SC, $P(t_{\rm al,1}< t_{\rm in})=0.92$ for LE and $P(t_{\rm al,1}< t_{\rm in})=0.93$ for SE.   
Two main effects are here combined: while the E models present higher spin magnitudes (hence longer alignment times) than the C models, they also predict a steeper profile in the mass ratio (Fig.~\ref{qfit}), with fewer small-$q$ binaries (hence, on average, shorter alignment times). 
Those misaligned, rapidly rotating, primaries BHs predicted by the E  models are ideal targets for strong-gravity precession effects in the late inspiral and merger (\citealt{schnittman04,kesdengerosa15}; Sec.~\ref{kicks}). 

Fig.~\ref{cumdist} also provides intuitions on the consequences of systematic errors in our time-scale estimates. If the inspiral (alignment) time is 
larger (smaller) of a factor $\tau=10$, all binaries in the sample align by the end of the gas-driven inspiral. On the other hand, if the inspiral (alignment) process is $10$ times faster (slower), i.e. $\tau=0.1$, only 60\%-80\%  of the primaries aligns.

\subsection{Differential misalignment and  kick velocity}
\label{kicks}

The most notable consequence of our findings is a clear prediction for the spin-orientation  angles at the onset of the GW-driven inspirals: a  non-negligible fraction of supermassive BH binaries  approaches the GW-driven phase with  $\theta_1 \neq 0$ and $\theta_2\simeq 0$. 

 If the binary lies in the same plane of the circumbinary disc  (\citealt{ivanov99,miller13}; see Sec.~\ref{discussion}), the angles $\theta_i$ may  be taken as estimates of the misalignment between the BH spins and the binary angular momentum $\mathbf{L}_{\rm bin}$, and used to estimate the properties of the post-merger BH. While final mass \citep{bmr12} and spin \citep{barrez09} do not critically depend on the spin misalignments, these are crucial to predict the final recoil \citep{campanelli2007b,gonzalez2007}. The largest kick velocities 
(up to $\sim 5000$ Km s$^{-1}$) are attained for maximally spinning, equal-mass BH mergers with moderately large misalignments $\theta_i\sim 50^\circ$ \citep{lousto11,lousto13}.  

Here we perform  a preliminary study to estimate the impact  of our findings on the kick velocity distribution. To maximize the effect, we consider maximally spinning BHs in  binaries with mass ratio $q=0.2$, right at the onset of the misaligned regime highlighted in Sec.~\ref{fixedparameters} (cf. Fig.~\ref{band}). 
Numerical-relativity fitting formulae are available to compute kick velocities, but precession effects  during the GW-driven inspiral must be taken into account,  especially for  configurations with sensibly different spin tilts  $\theta_1 \neq \theta_2$ \citep{schnittman04,kesden2010a,kesden2010b,kesdengerosa15,berti12,gerosa13}.  GWs start driving the merger at the decoupling radius  (\citealt{armitagenararajan02,gold14}; see also \citealt{milosavljevic05})
\begin{align}
r_{\rm dec} \simeq 760 \frac{G M_{\rm bin}}{c^2} \left(\frac{\alpha}{0.2}\right)^{-2/5}  \left(\frac{H/r}{0.001}\right)^{-4/5} \frac{(7.2 \,q)^{2/5}}{(1+q)^{4/5}} \;,
\end{align}
where the angular momentum losses in GWs dominate over the viscous evolution of the disc. We first transfer the spin orientations from the initial separation $r_{\rm dec} = 760\,G M_{\rm bin}/ c^2$ to $r_{\rm fin}=10 \,G M_{\rm bin}/ c^2$ using the precession-averaged formalism recently presented by  \cite{kesdengerosa15}; and we finally apply\footnote{We refer the reader to Sec.~2.1 in \cite{gerosa15} for details on the fitting formulae implementation. In their notation, we assume a random initial phase $\Delta\Phi$ at $r_{\rm dec}$. 
In order to disentangle the dependence of the kick velocity on the spin orientations, we maximize over the orbital phase at merger $\Theta$ (thus only showing the maximum kick allowed in each configuration). The relevance of $\Theta$ on the results presented in this section can be easily predicted because the kick velocity scales roughly as $v_k\propto \cos\Theta$ [c.f. e.g. Eq.~(2) in \cite{campanelli2007b}]. The kick velocity is independent of $M_{\rm bin}$.} the numerical-relativity fitting formula  by \cite{lousto13} at $r_{\rm fin}$.

\begin{figure}\center
\includegraphics[width=\columnwidth]{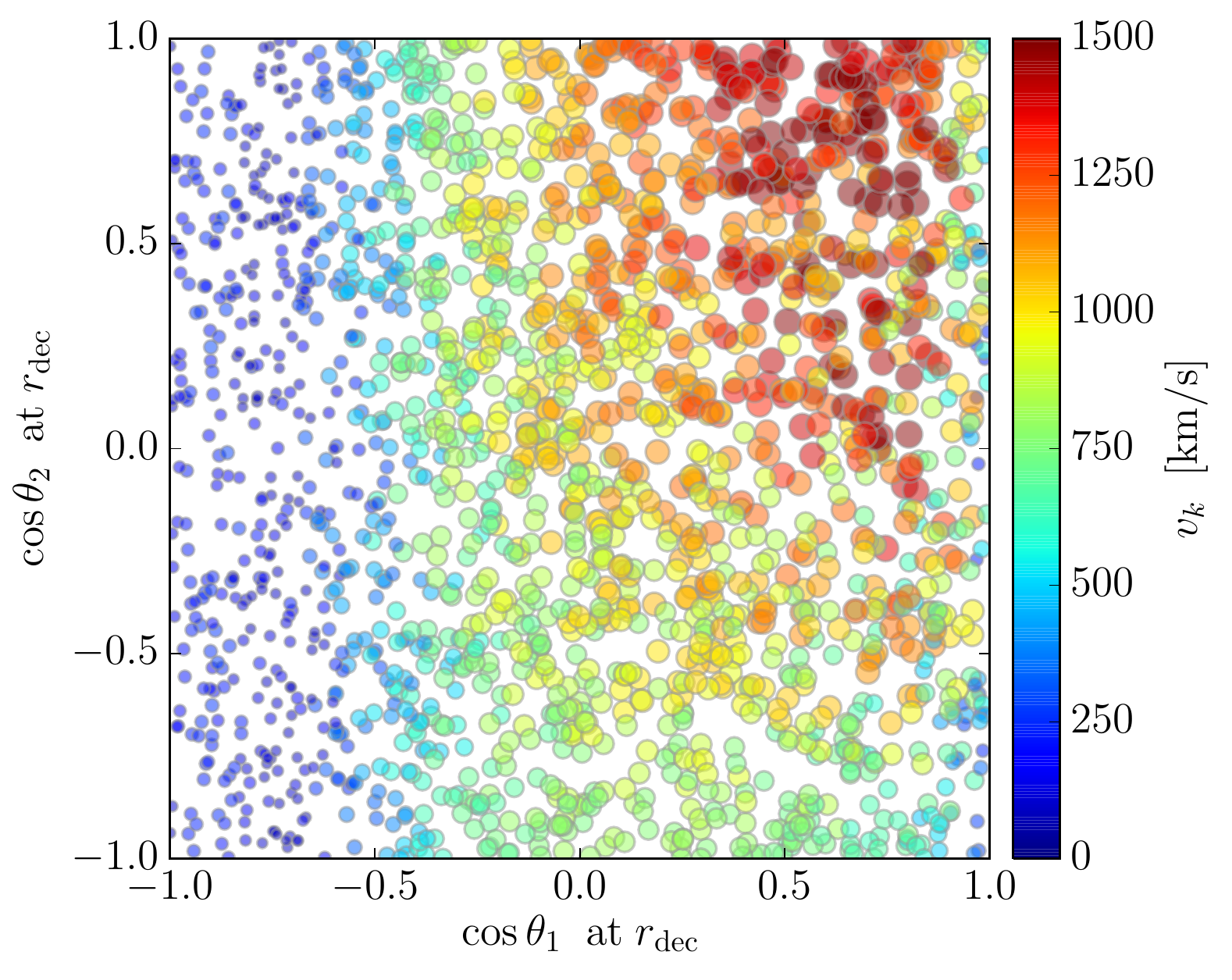}
\caption{
Maximum kicks velocity $v_{\rm kick}$ (colour scale and marker size) as a function of the misalignment angles $\cos\theta_i= \mathbf{S_i} \cdot \mathbf{L_{\rm bin}}$, measured at the decoupling radius $r_{\rm dec}$ ($x$ and $y$ axes for primary and secondary BHs respectively). We consider maximally spinning BH binaries with $q=0.2$, predicted to be at the onset of the misaligned regime unveiled by our astrophysical model. Large kicks are foreseen in the $\theta_1 \neq 0$, $\theta_2\simeq 0$ region, predicted to be astrophysically relevant. Superkicks with $v_{k}\gtrsim 2000$ km/s are not likely in gas-rich environments because binaries with larger mass ratio are expected to align before mergers.}
\label{t1t2kick}
\end{figure}

Fig.~\ref{t1t2kick} relates the spin orientation at the decoupling  radius to the maximum kicks velocities allowed in each configuration. Notably, higher kicks are found in the $\theta_1 \neq 0$, $\theta_2\simeq 0$ region,  which we predict to be populated by the Bardeen-Petterson effect ($\sim 8\%$ of the cases from the models used in Sec.~\ref{cosmological}). The evolution of the spin orientation in the GW-driven inspiral can be qualitatively understood in terms of two special families of configurations (\emph{spin-orbit resonances}, \citealt{schnittman04}), in which the projections of the two spins on the orbital plane are either aligned or anti-aligned to each other. Kicks are  suppressed (enhanced) by spin-precession effects for binaries in the aligned (anti-aligned) family \citep{kesden2010b}. Configurations lying in the $\theta_1 \neq 0$, $\theta_2\simeq 0$ region of the parameter space are likely to be attracted into the anti-aligned family (cf. e.g. Fig.~5 in \citealt{gerosa13}): as binaries approach the merger phase, most of their spin-precession cycles is spent with the two spins forming an angle $\sim \pi$ when projected on to the orbital plane. These configurations are qualitatively similar to the standard \emph{superkick} configuration \citep{campanelli2007b,gonzalez2007} and notably predict high kick velocities.

Our prediction is that, if merging rapidly rotating BHs are present in gas-rich environments, kicks as large as $v_k\sim 1500$ km/s can happen. Such kicks can make the BH wander in the galaxy outskirt for times as long as $10- 100$ Myr with displacements of $\sim 10^3$pc \citep{gualandris08,komossa08b,sijacki11,gerosa15}, possibly at the level of observational consequences \citep{komossa12}. Larger values of $v_k$ are only  possible in merging binaries with mass ratio closer to the equal-mass case (for more quantitative information see e.g. Fig.~3 in \citealt{lousto13}). Both BHs in these binaries are predicted to be found aligned at merger $\theta_1\sim\theta_2\sim 0$  (Sec.~\ref{fixedparameters}), which limits the kick velocity to $\sim 300$ km/s. Our analysis shows that \emph{superkicks} with $v_k \gtrsim 2000$ km/s are disfavoured in gas-rich environments where the Bardeen-Petterson effect comes into play. 

\section{Discussion and conclusions}
\label{discussion}

Alignment of BH spins in merging BH binaries may be differential. Using a semi-analytical model, we find that light secondaries may accrete almost all mass leaking tough the binary gap and prevent primary BHs from alignment. In particular, such differential alignment occurs for binary with mass ratio $q \lesssim 0.2$. Gaseous interactions have enough time to align both spins in binaries with mass ratio closer to the equal-mass case (Sec.~\ref{fixedparameters}). We implement our analysis trough a time-scale argument, comparing the time needed to align the BH spins in the Bardeen-Petterson effect $t_{\rm al}$ to the total time available in the gas-driven inspiral phase $t_{\rm in}$. The alignment and the inspiral processes are coupled by the accretion rates: while the binary migration is set by the circumbinary disc mass rate, alignment is powered by the mass accreting on to each BH. Mass from the circumbinary disc is expected to pile up at the outer edge of the cleared cavity, suppressing the alignment process. On top of this, mass leaking trough the cavity is found to preferentially accrete on to the secondary BH which orbits closer to the disc edge. This causes the alignment time of the primary BH to be several orders of magnitudes longer than that of the secondary, and
possibly even longer than the inspiral time. Differential accretion is a key, previously neglected, feature to tackle the spin-alignment problem: for comparison, \cite{miller13} only quoted a factor of $\sim q^{-1/2}$ between the alignment times of the two BHs.   While powerful for its simplicity, our time-scale argument fails to capture the dynamics of the alignment process: more elaborate models involving numerical simulations are needed to predict the residual misalignment 
of primary BHs that cannot be aligned with the Bardeen-Petterson effect, and to estimate how close to complete alignment 
secondaries can be found in realistic environments.

We present preliminary results to address the relevance of our findings on to the supermassive-BH cosmic history. Using publicly available synthetic populations, we find that binaries in differential misalignment are expected in realistic cosmological scenarios (Sec.~\ref{cosmological}). A fraction of  $\sim 8 \%$ of the BH primaries are found misaligned at merger even in models predicting large spin magnitudes,  opening for the possibility of large kick velocities. Merging BHs with spin angles $\theta_2\sim 0$ and $\theta_1\neq 0$ are subject to the largest kicks velocities available for their mass ratio and spin magnitudes. In particular, misaligned primaries in BH binaries with $q \simeq 0.2$ may suffer kicks as large as $\sim 1500$ km/s, while higher mass ratios are needed to succeed in achieving proper \emph{superkicks}   (Sec.~\ref{kicks}).  
Binaries approaching the merger phase with differentially misaligned spins will exhibit pronounced precession effects in the later GW-driven inspiral phase \citep{schnittman04,kesdengerosa15}. Orbital plane precession  modulates the amplitude of the GW cycles, encoding information of the astrophysical environment on to the emitted GW pattern (\citealt{gerosa14,vitale14}; Trifir\`o et al. in preparation). These features may in principle be used to  recall and constrain our models using future space-based GW observations, although more work is needed to quantify these statements. 

Several assumptions have been made in developing our models, some of them worth of future improvements. First and perhaps most importantly, our model estimates whether spin misalignments, \emph{if  present}, are carried over towards merger. The same dynamical processes that bring the binary together may play a role in determining the spin directions before  Type II migration takes place. While star scattering is unlikely to affect the spin orientations because it does not present any preferred direction, this may not be the case for previous larger-scale gas interactions. Dynamical friction against gaseous environment may be crucial to promote the binary formation: even if short ($\sim 10$ Myr;  \citealt{escala05,dotti07}), this phase presents interesting dynamics involving tidal shocks and nuclear cusp disruption \citep{vanwass14} whose possible consequences on the spin directions still need to be explored.  
Secondly, our model only estimates whether the BH spins aligns to the angular momentum of the disc, while strong-gravity effects in latest inspiral and merger phase depend on the misalignments between the BH spins and the binary angular momentum. The further assumption of alignment between the binary orbital plane and the circumbinary disc  \citep{ivanov99} is necessary to estimate the properties of the post-merger BHs, in particular the kick velocity.
Thirdly, we have neglected the BH mass growth during the alignment process. Differential accretion brings binary towards larger mass ratios on time-scales $\sim M/ \dot M$. While this effect can be safely neglected on the time-scale of the alignment process $t_{\rm al} \sim 10^{-3} M/ \dot M$ [cf. Eq.~(\ref{talign})], it  may not be negligible on the time-scale of the inspiral. However, this point may only be important for aligned binaries which do not present large kick velocities anyway. As extensively discussed is Sec.~\ref{companion}, we also neglect the presence of the companion when estimating the alignment time (cf. \citealt{miller13}).  However, Fig.~\ref{comp_contour} shows that this effect (a factor of $\sim 2$ in $t_{\rm al}$) mostly affects $q \sim 1$ binaries where both BHs aligns anyway. This point is worth  further investigation, but sensible modelling efforts are likely to be required because the presence of two external torques (Lense-Thirring precession and the companion) cannot be fully captured within a time-scale argument \citep{martin09}.   Finally, we have assumed that all values of $q$ are allowed on cosmological grounds. 
Assuming the BH mass correlates with the galaxy mass, galaxy pairs with $q \lesssim 0.1$ may fail in forming close binaries because of strong tidal interactions before the galactic merger \citep{taffoni03,callegari09,vanwass14}. Mass stripped away from the secondary galaxy may sensibly increase the delay time between the galaxy and the BH mergers, possibly even preventing the BH binary formation.

We stress that the impact of the disc aspect ratio on the Bardeen-Petterson effect is still not understood and can potentially be crucial. Both  migration process (Sec.~\ref{gasdriventime}) and  gas streaming through the binary cavity (Sec.~\ref{diffacc})  have only been simulated with sensibly thicker disc (typically $H/r\sim 0.01 - 0.1$) than those predicted for discs surrounding supermassive BHs ($H/r\sim 0.001$, see Sec.~\ref{fragmentation}). In particular, a sensibly lower amount of gas may be able to leak trough the cavity in thinner discs, possibly slowing down the alignment process.  
 Although we are aware of the computational constraints in simulating thin discs, we stress that such simulations are needed to validate the analytical expressions assumed here, and we point towards the importance of pushing these numerical efforts to lower values of the aspect ratio.

Merging supermassive BH binaries are unique systems where gravity and astrophysics both play together to shape the dynamics.
BH spin alignment (or misalignment) is an imprint of angular momentum transfer between the astrophysical and the relativistic side of BH binaries whose potential still need to be fully uncovered.

\section*{Acknowledgements}
We especially thank Gordon Ogilvie for providing us the code to compute the warp-propagation coefficient $\alpha_2$,  Brian Farris for the data that enter in Fig.~\ref{twofit}, Cole Miller for insightful discussion on the role of the companion-induced precession and Alberto Sesana for help with the synthetic binary populations. We thank Cathie Clarke and the participants of the 2014 Milan Christmas workshop for stimulating discussions on our preliminary results; together with Michael Kesden, Emanuele Berti, Ulrich Sperhake, Richard O'Shaughnessy and the gravitational-physics group at the University of Mississippi for early discussions on the inspiral-time-scale interpolation and the kick predictions.
DG is supported by the UK Science and Technology Facility Council and the Isaac Newton Studentship of the University of Cambridge; partial support is also acknowledged from FP7-PEOPLE-2011-CIG Grant
No. 293412, FP7-PEOPLE-2011-IRSES Grant No.295189, SDSC and TACC
through XSEDE Grant No.~PHY-090003 by the NSF, Finis Terrae through
Grant No.~ICTS-CESGA-249, ERC-2013-ADG Grant No. 341137, STFC Roller Grant No. ST/L000636/1 and
DiRAC's Cosmos Shared Memory system through BIS Grant No.~ST/J005673/1
and STFC Grant Nos.~ST/H008586/1, ST/K00333X/1.
 Figures have been generated using the \textsc{Python}-based
\textsc{matplotlib} package \citep{python}.

\bibliography{lodato}

\end{document}